\documentclass[a4paper,fleqn]{cas-sc}

\usepackage[numbers,sort&compress]{natbib}
\usepackage[T1]{fontenc}
\usepackage{graphicx}
\usepackage{hyperref}
\usepackage{lastpage}
\AtBeginDocument{}
\makeatletter
\@newl@bel{r}{tab:data_desc}{{S1}{2}{Per-exchange data coverage, and sample sizes retained for the primary signed one-minute source-price reconstruction shared by DFA-2, MF-DFA, CECP, and DHVG. A segment is a maximal run within which no source-price gap of 15 minutes or longer occurs. Shorter gaps (2--14 minutes) are filled at zero return and do not start a new segment. The post-2020 production window used throughout the main text begins after all three cryptocurrency assets are simultaneously covered across this exchange universe (Sec.~\ref {sec:supp_scope})}{table.caption.1}{}}
\@newl@bel{r}{fig:supp_scope_rolling}{{S1}{6}{Rolling-window scaling diagnostics on deseasonalized minute returns, 2015--2025 ($W = 20{,}000$ minutes, centered moving average of length $30$). (a) Local DFA-2 Hurst exponent of returns (dashed line: random-walk benchmark $H = 0.5$). (b) Local volatility-ACF decay exponent. Faint and dark traces show unsmoothed estimates and centered moving averages, respectively. In both panels, cryptocurrency trajectories converge to the equity benchmark around 2019--2020, thereafter evolving within a common dispersion band}{figure.caption.3}{}}
\@newl@bel{r}{fig:supp_scope_garch}{{S2}{6}{Rolling-window GARCH$(1,1)$ parameter distance from the S\&P~500 on deseasonalized minute returns, 2015--2025 ($W = 20{,}000$ minutes, moving average of length $30$). Faint traces show unsmoothed estimates. (a) Shock-sensitivity distance $|\alpha _1^{\text {crypto}} - \alpha _1^{\text {S\&P~500}}|$. (b) Persistence distance $|\beta _1^{\text {crypto}} - \beta _1^{\text {S\&P~500}}|$. Vertical dotted line marks January~1, 2020. Both distances drop to $\lesssim 0.05$ from 2020 onward, providing a parametric counterpart to the scaling diagnostics in Fig.~\ref {fig:supp_scope_rolling}}{figure.caption.4}{}}
\@newl@bel{r}{sec:supp_scope}{{S6}{5}{Post-2020 Cutoff Justification}{section.6}{}}
\@newl@bel{r}{sec:supp_cecp}{{S3}{2}{Complexity--Entropy Causality Plane}{section.3}{}}
\@newl@bel{r}{sec:supp_hvg_degree}{{S9}{10}{Degree-Resolved HVG Irreversibility}{section.9}{}}
\@newl@bel{r}{fig:supp_hvg_transform_grid}{{S7}{11}{HVG irreversibility across return transformations in the post-2020 window. Rows correspond to the absolute-return series, positive-return subsequence, and negative-return-magnitude subsequence. Each transform is applied within the source-price segments, and graph degree counts are pooled only after segment-wise construction. The first two columns report the Jensen--Shannon divergence $D_\mathrm {JS}$ and signed directional asymmetry $A_\mathrm {KL}$ for all degrees and for the high-degree sector $k\geq 10$. Surrogate-ensemble means are overlaid with $\pm $ s.d.\ error bars. The right column shows the local degree profile $A(k)$. Filled markers denote $n_k\geq 100$, open markers denote $20\leq n_k<100$, and crosses connected by dashed lines denote $n_k<20$. All supported natural-number degrees are connected to show the profile without implying equal precision in its sparse tail}{figure.caption.10}{}}
\@newl@bel{r}{sec:supp_hvg}{{S4}{3}{Directed Horizontal Visibility Graph Analysis}{section.4}{}}
\@newl@bel{r}{sec:supp_surrogates}{{S5}{4}{Surrogate Data Generation}{section.5}{}}
\@newl@bel{r}{tab:supp_tail_post2022}{{S2}{7}{Post-2022 deseasonalized upper-tail power-law slope estimates. The estimates use the same KS-selected $x_{\min }$ maximum-likelihood procedure as the main-text tail analysis}{table.caption.5}{}}
\@newl@bel{r}{tab:price_contiguous_k_grid}{{S3}{12}{Threshold and pseudocount sensitivity of the primary price-derived signed-return diagnostics over $K=8,\ldots,12$ and pseudocounts $0.1$ and $0.5$}{table.caption.11}{}}
\@newl@bel{r}{tab:sampling_interval_combined}{{S4}{15}{Combined sampling-interval sensitivity of the high-degree source-gap DHVG and CECP diagnostics}{table.caption.14}{}}
\@newl@bel{r}{fig:supp_mfdfa}{{S3}{8}{Robustness of the MF-DFA multifractal spectrum width $\Delta \alpha $ to the start-date cutoff. Solid bars show the post-2020 window and hatched bars show the post-2022 window. Surrogate ensemble results are overlaid as markers with error bars (plus markers for Shuffle and cross markers for IAAFT, showing mean $\pm $ s.d.\ over 100 realizations). Bitcoin retains the broadest spectrum in both windows, but the ordering among ETH, XRP, and S\&P~500 is not preserved: Ethereum's spectrum narrows the most, so XRP becomes the second-widest in the post-2022 window. The shuffle/IAAFT equivalence is preserved across both windows}{figure.caption.6}{}}
\@newl@bel{r}{fig:supp_cecp_grid}{{S6}{10}{Robustness of the CECP crypto--equity separation to embedding dimension $m$ and delay $\tau $, shown as grouped bars for each $m$ as a function of delay $\tau $. Left column: statistical-complexity ratio $C_{\mathrm {S\&P}}/\max (C_{\mathrm {crypto}})$. Right column: entropy-deficit ratio $(1-H)_{\mathrm {S\&P}}/\max (1-H)_{\mathrm {crypto}}$. Top row: post-2020 production window. Bottom row: post-2022 robustness window. All bars exceed one, indicating that the S\&P~500 remains lower-entropy and higher-complexity than all three cryptocurrencies throughout the tested embedding grid}{figure.caption.9}{}}
\@newl@bel{r}{fig:supp_cecp}{{S4}{8}{Robustness of the complexity--entropy causality plane (CECP) coordinates to the start-date cutoff. The figure compares the original and surrogate coordinates ($1-H$ in the left panel, $C$ in the right panel) between the post-2020 (solid bars) and post-2022 (hatched bars) windows. Original returns are plotted as bars, while surrogate ensemble results are overlaid as markers with error bars (cross markers for IAAFT and plus markers for Shuffle, showing mean $\pm $ s.d.\ over 100 realizations). All values are scaled by $10^4$. The cross-asset hierarchy and the relationships between the empirical data and the surrogates are consistent across both analysis windows}{figure.caption.7}{}}
\@newl@bel{r}{fig:supp_hvg}{{S5}{9}{Comparison of signed-return HVG directional-asymmetry diagnostics between the post-2020 and post-2022 windows. Diagnostics are computed over all degrees for (a) $D_{\mathrm {JS}}$ (log scale) and (b) $A_{\mathrm {KL}}$ (linear scale), and restricted to the high-degree sector $k\geq 10$ in (d, e). Panel (c) shows the degree-wise asymmetry profile $A(k)$ for each asset separately, with the post-2020 window as a solid line and the post-2022 window as a dashed line; the gray band marks the $k\geq 10$ sector used in (d, e), and all four subpanels share a common $A(k)$ axis. For panels (a, b, d, e), the post-2020 window is shown as solid (plain) bars and the post-2022 window as hatched bars. Marker style encodes the per-degree support $n_k=c_\mathrm {in}(k)+c_\mathrm {out}(k)$, independent of window: filled markers denote $n_k\geq 100$, open markers $20\leq n_k<100$, and $\times $ markers $n_k<20$. In (a, b, d, e), open crosses (IAAFT) and plus markers (Shuffle) indicate surrogate ensemble means $\pm $ one standard deviation}{figure.caption.8}{}}
\@newl@bel{r}{sec:supp_data_construction}{{S2}{1}{Data Construction}{section.2}{}}
\@newl@bel{r}{subsec:supp_cecp_sampling}{{S10.2}{14}{Complexity--Entropy Causality Plane}{subsection.10.2}{}}
\@newl@bel{r}{sec:supp_post2022}{{S7}{7}{Post-2022 Re-analysis}{section.7}{}}
\@newl@bel{r}{eq:supp_perm_prob}{{S3}{2}{Probability Distribution of Ordinal Patterns}{equation.3.3}{}}
\makeatother

\usepackage{bm}
\usepackage{kotex}
\usepackage{amsmath}
\usepackage{amssymb}
\usepackage{booktabs}
\usepackage{multirow}
\usepackage{makecell}
\usepackage{tabularx}
\usepackage{subcaption}
\usepackage{enumitem}
\usepackage{tikz}
\usetikzlibrary{arrows.meta}
\newlist{cellitem}{itemize}{1}
\setlist[cellitem]{nosep, leftmargin=1em, label=\textbullet, topsep=0pt, partopsep=0pt, before*={\vspace{-2pt}}, after*={\vspace{-2pt}}}

\begin{document}
\let\WriteBookmarks\relax
\shorttitle{Universality and Heterogeneity of Stylized Facts in Cryptocurrency and Equity Markets}
\shortauthors{Kim et al.}

\title [mode = title]{Universality and Heterogeneity of Stylized Facts in Cryptocurrency and Equity Markets}

\author[1]{Jaesung Kim}[orcid=0009-0008-7268-8576]

\author[1]{Changhee Cho}

\author[1,2,3]{Jae Woo Lee}
\cormark[1]
\ead{jaewlee@inha.ac.kr}

\cortext[1]{Corresponding author}

\affiliation[1]{organization={Department of Physics, Inha University},
    city={Incheon},
    postcode={22212},
    country={Republic of Korea}}

\affiliation[2]{organization={Institute of Quantum Science, Inha University},
    city={Incheon},
    postcode={22212},
    country={Republic of Korea}}

\affiliation[3]{organization={School of Computational Sciences, Korea Institute for Advanced Study},
    city={Seoul},
    postcode={02455},
    country={Republic of Korea}}
\begin{abstract}
    This study investigates whether the macroscopic statistical maturity of cryptocurrencies implies dynamical equivalence with an equity-market benchmark.
    We analyze one-minute data (2020--2025) for Bitcoin, Ethereum, XRP, and E-mini S\&P~500 futures using the Complexity--Entropy Causality Plane (CECP) and directed horizontal visibility graphs (DHVG).
    While conventional stylized facts overlap across the analyzed assets, cryptocurrencies exhibit weaker local ordinal organization and a larger divergence between their high-degree in- and out-visibility distributions relative to matched surrogate baselines under the primary specification.
    This high-degree separation is strongest at the one-minute scale and weakens under 5- and 10-minute aggregation, while the direction of the underlying asymmetry remains more asset- and preprocessing-dependent.
    We therefore conclude only that macroscopic similarity can coexist with diagnostic-specific structural differences in the assets and period examined.
\end{abstract}


\maketitle

\section{Introduction}

The approval of a spot Bitcoin Exchange-Traded Fund (ETF) by the U.S. Securities and Exchange Commission (SEC) in 2024, together with the rapid growth of market capitalization and institutional participation, indicates that cryptocurrencies are no longer peripheral speculative instruments.
Major cryptocurrencies now interact with conventional financial institutions, derivatives markets, and systematic trading strategies, while still operating through fragmented exchanges and continuous 24/7 trading.
This combination makes them a useful empirical setting for asking whether a market can become statistically mature without becoming dynamically equivalent to traditional financial markets.

Classical models for Value-at-Risk (VaR), option pricing, and portfolio allocation often rely on idealized assumptions such as Gaussian returns or independent and identically distributed increments.
Empirical studies in econophysics and financial econometrics have repeatedly shown that conventional markets violate these assumptions through heavy-tailed returns, volatility clustering, long memory in activity, and multiscale fluctuations, collectively known as stylized facts~\cite{Mantegna1999, Cont2001, ANDERSEN1997115, DACOROGNA2001121, PhysRevE.60.1390}.
These regularities provide model-independent benchmarks for market comparison, because they summarize features directly relevant to tail risk, market efficiency, volatility persistence, and liquidity formation~\cite{Mandelbrot1963, Fama1970}.

For cryptocurrency markets, the central question has shifted from whether such stylized facts exist to what their apparent convergence with conventional markets actually means.
If cryptocurrencies and equity-index futures share heavy tails, near-random-walk return scaling, and persistent absolute returns, their shared statistical properties may suggest that standard trading and pricing models can be applied interchangeably to both markets.
However, two return series can display similar distributional and scaling behavior while differing in the temporal organization that produces that behavior.
Risk models based only on tail slopes, Hurst exponents, or multifractal widths may therefore miss structural differences associated with nonlinear temporal ordering, local ordinal constraints, or time-directed asymmetry.

Prior research on cryptocurrency stylized facts initially focused on Bitcoin (BTC), documenting power-law-like tails, non-Gaussian return distributions, and long-range memory~\cite{BEGUSIC2018400, Drozdz2018, Urquhart2016, Bariviera2017, Kristoufek2018}.
Broader evidence shows that return-tail behavior also varies with cryptocurrency age and market capitalization~\cite{Pessa2023}.
Later studies identified multifractality in BTC returns~\cite{Takaishi2021, Lahmiri2019}, examined broader cryptocurrency panels, and showed that part of the observed multifractality may arise from broad marginal distributions rather than nonlinear dynamics alone~\cite{STOSIC201954}.
High-frequency evidence suggests that some cryptocurrency markets have matured toward statistical patterns seen in established foreign-exchange markets~\cite{WATOREK20211}, while ordinal studies show that efficiency varies across time and assets in both cryptocurrency and stock markets~\cite{Sigaki2019,Alves2020}.
Related work has also represented cryptocurrency price fluctuations using ordinal-partition networks~\cite{Shahriari2022}.
Nevertheless, a direct comparison with a conventional equity benchmark requires more than documenting the presence of stylized facts.
It must separate marginal-distribution effects, linear-correlation effects, and residual temporal structure, especially because intraday seasonality in equity-index futures can otherwise create apparent complexity that is not intrinsic to the return dynamics.

This study addresses this issue by treating stylized facts as a starting point rather than as the conclusion of the comparison.
We analyze one-minute 2020--2025 data for BTC, Ethereum (ETH), Ripple (XRP), and E-mini S\&P~500 futures.
To reduce reliance on individual trading venues, we construct volume-weighted composite cryptocurrency prices, calculate their log-returns, and apply intraday deseasonalization to all assets. These procedures do not eliminate differences in trading calendars or market microstructure.
We then implement a three-layered diagnostic framework to disentangle apparent convergence from structural heterogeneity.
First, macroscopic stylized facts, including heavy-tail CCDF and DFA scaling, establish the distributional and correlational properties.
Second, multifractal analysis using MF-DFA and surrogate tests identifies the residual scaling heterogeneity that persists beyond linear correlations.
Third, structural diagnostics using the complexity--entropy causality plane (CECP) and directed horizontal visibility graph (directed HVG) identify residual heterogeneity in local ordinal patterns and time-directed asymmetry.

The present study therefore seeks to address the following two central questions.
\begin{enumerate}
    \item Do the selected cryptocurrencies and the equity-index futures benchmark exhibit similar distributional and scaling properties during the study period?
    \item Where such similarities occur, do CECP and directed-HVG diagnostics nevertheless reveal differences in ordinal organization and time-directed visibility structure?
\end{enumerate}

Our results support a layer-dependent interpretation: coarse-grained similarity coexists with structural heterogeneity.
At the macroscopic level, all four assets share heavy-tailed distributions, return Hurst exponents close to the random-walk benchmark, persistent absolute-return fluctuations, and broadly overlapping MF-DFA curves.
The original--surrogate multifractal-width residual remains positive, but it varies across individual assets rather than producing a simple crypto--equity class separation.
Structural diagnostics, however, show specification-dependent differences in ordinal organization and time-directed visibility, most clearly in high-degree $D_\mathrm{JS}$ under the primary one-minute analysis.
The comparison is limited to the selected assets, the 2020--2025 period, and the stated preprocessing and sampling specifications.

The remainder of this paper is organized as follows.
Section~\ref{sec:data_method} describes the data window, the intraday-deseasonalization protocol, and the analytical methodology.
Section~\ref{sec:global} establishes the standard stylized facts and their limits as a structural classifier.
Section~\ref{sec:structural_diagnostics} then applies surrogate-controlled CECP and directed-HVG diagnostics to test residual crypto--equity heterogeneity.
Section~\ref{sec:conclusion} discusses the implication that macroscopic similarity does not entail structural equivalence and concludes the paper.

\section{Data and Methodology}
\label{sec:data_method}

\subsection{Data Description}
\label{subsec:data_desc}

We analyze high-frequency trading data for three major cryptocurrencies---Bitcoin (BTC), Ethereum (ETH), and Ripple (XRP)---sourced from major global USD exchanges (Bitstamp, Coinbase, and Kraken), with per-exchange data coverage summarized in Table~\ref*{tab:data_desc} of the Supplementary Material.
The E-mini S\&P~500 futures contract (CME) serves as the conventional equity benchmark (hereafter referred to simply as the S\&P~500).
Its broad US equity-index exposure, liquid electronic trading, and near-continuous coverage during the futures week make it a useful high-frequency counterpart to cryptoassets that trade continuously around the clock.
It remains a single futures benchmark, with scheduled trading breaks and holiday closures that distinguish it from 24/7 cryptocurrency markets.
Comparative studies of stock-market indices, including the S\&P~500, NASDAQ~100, and STOXX~50, report similar long-memory and volatility-clustering behavior across U.S. and European markets~\cite{Bentes2008}.
Heavy tails and volatility persistence are well-established features of equity and high-frequency financial markets~\cite{Cont2001,ANDERSEN1997115,DACOROGNA2001121}; the corresponding stylized-fact baseline for the E-mini series is documented in Fig.~\ref{fig:ccdf} and Table~\ref{tab:hurst}.
The analysis period for all assets spans from January 1, 2020, to December 31, 2025, an operational comparison window motivated by the rolling diagnostics below.
Rolling-window DFA, volatility-autocorrelation, and GARCH(1,1) parameter-distance diagnostics computed over 2015--2025 show that BTC, ETH, and XRP converge onto the S\&P~500's dispersion band around 2019--2020, with XRP the slowest asset to transition (Figs.~\ref*{fig:supp_scope_rolling}--\ref*{fig:supp_scope_garch}; Sec.~\ref*{sec:supp_scope} of the Supplementary Material for the full diagnostic protocol).

Let $P_t$ represent the price of a cryptocurrency at time $t$.
The log-return over a 1-min interval is defined as $r_t = \ln (P_t / P_{t-1})$.
For cryptocurrencies, the preprocessing constructs a volume-weighted composite price across the available venues:
\begin{equation}
    \bar{P}_t = \frac{\sum_{i=1}^{N} V_{t,i} P_{t,i}}{\sum_{i=1}^{N} V_{t,i}},
    \qquad r_t=\ln\left(\frac{\bar P_t}{\bar P_{t-1}}\right),
    \label{eq:exchange_avg}
\end{equation}
where $P_{t,i}$ and $V_{t,i}$ denote the price and trading volume of venue $i$. Thus the analysis uses returns of the composite price, not a weighted average of venue-level returns.
The absolute return $|r_t|$ is used as the proxy for volatility.

Non-finite observations and non-trading intervals are not filled or bridged.
Cryptocurrency weekends are retained where available, whereas the futures series follows its trading calendar.
Details of the retained samples, timestamp-gap handling, and continuous-futures construction and rollover adjustment~\cite{MasteikaSaulius2012Cfds} are provided in Sec.~\ref*{sec:supp_data_construction} of the Supplementary Material.
To ensure a robust comparison between 24/7 cryptocurrency trading and the session-based equity market, we applied a minute-of-day volatility deseasonalization to all assets~\cite{ANDERSEN1997115}:
\begin{equation}
    \tilde{r}_t = \frac{r_t}{\hat{\sigma}_{m(t)}},
    \label{eq:deseas}
\end{equation}
where $m(t)$ is the US-Eastern minute of day and $\hat{\sigma}_{m}$ is the Gaussian-smoothed (width 30 min) standard deviation of signed returns. Below, $r_t$ denotes $\tilde r_t$ unless otherwise stated. Trading volume $V_t$ is analyzed separately.
All return-based analyses below therefore use the deseasonalized returns $\tilde r_t$ defined in Eq.~\ref{eq:deseas} (Eq.~2), unless otherwise stated.

\subsection{Heavy-Tail Analysis}
\label{subsec:stat_dist}

To establish the shared distributional properties, we characterize the upper tail of absolute returns using the complementary cumulative distribution function (CCDF):
\begin{equation}
    P(|\tilde{r}| > x) \sim x^{-\alpha},
    \label{eq:ccdf}
\end{equation}
where $\alpha$ denotes the CCDF tail exponent; the corresponding density exponent is $\alpha+1$.
The exponent $\alpha$ is estimated via maximum likelihood following the Clauset--Shalizi--Newman procedure~\cite{Clauset_2009}.
The lower threshold $x_{\min}$ is determined by minimizing the Kolmogorov--Smirnov distance between the empirical and fitted distributions.
These estimates provide a reference for macroscopic tail risk similarity rather than a definitive ranking of market convergence.

\subsection{Detrended Fluctuation and Multifractal Scaling Analysis}
\label{subsec:dfa_mfdfa_method}

To quantify temporal correlations and scale-dependent heterogeneity while reducing the influence of local trends, we used detrended fluctuation analysis (DFA) and its multifractal extension (MF-DFA)~\cite{PhysRevE.49.1685, KANTELHARDT2001441, PhysRevE.65.041107, PhysRevE.64.011114, Kantelhardt_2002}.
DFA divides the integrated time series into windows of size $s$, removes polynomial trends within each window, and measures the fluctuation function $F(s)$.
For a self-similar process,
\begin{equation}
    F(s) \sim s^H,
    \label{eq:dfa_h}
\end{equation}
where the Hurst exponent $H \approx 0.5$ corresponds to an uncorrelated process, while $H>0.5$ and $H<0.5$ indicate persistent and anti-persistent correlations, respectively.

We applied DFA to returns $r_t$, absolute returns $|r_t|$, and trading volume $V_t$ to examine market efficiency, volatility clustering, and persistence in trading activity, respectively.
Throughout the DFA and MF-DFA analyses, we used second-order polynomial detrending over the scaling range $s = 10^1$--$10^4$ minutes.\footnote{Using linear detrending instead did not materially affect the main conclusions.}
To examine whether small and large fluctuations obey different scaling laws, we applied MF-DFA by varying the moment order $q$ over $[-4,4]$.
The generalized Hurst exponent $h(q)$ was estimated from $F_q(s) \sim s^{h(q)}$.
From the generalized Hurst exponent, the multifractal spectrum $f(\alpha)$ is obtained via a Legendre transform of the mass exponent $\tau(q) = qh(q) - 1$:
\begin{equation}
    \alpha = \frac{d\tau(q)}{dq}, \quad f(\alpha) = q\alpha - \tau(q),
    \label{eq:legendre}
\end{equation}
where $\alpha$ is the singularity strength.
The spectrum width $\Delta\alpha = \alpha_{\max} - \alpha_{\min}$ was used as a scalar measure of multifractality, with larger values indicating stronger heterogeneity in local scaling behaviors.

\subsection{Complexity--Entropy Causality Plane}
\label{subsec:cecp_method}

To characterize the ordinal information structure and nonlinear dynamics, we employed the complexity--entropy causality plane (CECP)~\cite{PhysRevLett.99.154102} based on Bandt--Pompe patterns~\cite{Bandt2002}.
This symbolic approach maps continuous returns onto a sequence of discrete permutations, effectively capturing temporal organization while remaining robust against heavy-tailed marginal distributions.

The ordinal patterns, defined by embedding dimension $D$ and delay $\tau$, yield a distribution $P$ of permutations.
The normalized permutation entropy $H[P]$, quantifying the degree of disorder, is:
\begin{equation}
    H[P] = \frac{-\sum_{\pi} p(\pi)\log p(\pi)}{\log(D!)},
    \label{eq:normalized_permutation_entropy}
\end{equation}
where $D!$ is the number of possible states and $p(\pi)$ is the probability of each possible permutation (Eq.~\ref*{eq:supp_perm_prob}).
The statistical complexity $C[P]$ measures non-trivial structural organization:
\begin{equation}
    C[P] = Q_J[P,P_e]\,H[P],
    \label{eq:statistical_complexity}
\end{equation}
where $P_e$ is the uniform distribution and $Q_J$ is the normalized Jensen--Shannon divergence.
$C[P]$ vanishes in the limits of pure randomness ($H \to 1$) and pure order ($H \to 0$), identifying structured dynamics between these poles.
We used $D=5$ and $\tau=1$, with $2$--$3\times10^6$ observations depending on asset (Table~\ref*{tab:data_desc}; Sec.~\ref*{sec:supp_cecp} of the Supplementary Material). Here $H[P]$ denotes permutation entropy, distinct from the DFA exponent $H$.

\subsection{Directed Horizontal Visibility Graph}
\label{subsec:hvg_method}

To quantify temporal asymmetry in market returns, we used the directed horizontal visibility graph (directed HVG)~\cite{Luque2009, Lacasa2012}.
This approach characterizes time-series irreversibility by comparing in-degree and out-degree distributions using information-theoretic divergences.
Previous studies demonstrated that financial irreversibility fluctuates over time, clusters during market stress, and provides information complementary to standard volatility~\cite{FLANAGAN20161689}.
In this study, we analyzed directed-HVG irreversibility by decomposing the in/out-degree asymmetry into three complementary components: (i) the magnitude of asymmetry, (ii) its directional orientation, and (iii) the degree-local concentration.
Ordinal-pattern frequencies can also test time irreversibility~\cite{Zanin2018}. We use DHVG because it localizes directional mismatch by node degree, whereas the CECP coordinates are not themselves an irreversibility test.

\begin{figure}
    \centering
    \resizebox{0.65\textwidth}{!}{\definecolor{axisgray}{HTML}{374151}
\definecolor{barfill}{HTML}{F3F4F6}
\definecolor{bardraw}{HTML}{4B5563}
\definecolor{focusfill}{HTML}{FDE7C3}
\definecolor{focusdraw}{HTML}{B45309}
\definecolor{inblue}{HTML}{2F6F9F}
\definecolor{outorange}{HTML}{D97706}
\definecolor{blockedgray}{HTML}{667085}

\begin{tikzpicture}[
    x=0.95cm,
    y=0.36cm,
    line cap=round,
    line join=round,
    >={Stealth[length=2.5mm, width=1.8mm]},
    font=\sffamily
]
    \def\w{0.15}

    \draw[axisgray, line width=0.9pt, ->]
        (-0.45,0) -- (10.05,0)
        node[right=1pt, text=axisgray] {$t$};
    \draw[axisgray, line width=0.9pt, ->]
        (0,-0.35) -- (0,9)
        node[above=1pt, text=axisgray] {$r(t)$};

    \begin{scope}[draw=inblue, line width=1.15pt, ->]
        \draw (1+\w,6) -- (4-\w,6);
        \draw (3+\w,3.5) -- (4-\w,3.5);
    \end{scope}

    \begin{scope}[draw=outorange, line width=1.15pt, ->]
        \draw (4+\w,7) -- (9-\w,7);
        \draw (4+\w,5) -- (8-\w,5);
        \draw (4+\w,3) -- (6-\w,3);
        \draw (4+\w,1) -- (5-\w,1);
    \end{scope}

    \node[text=inblue, above=2pt, inner sep=0pt] at (2.45,6)
        {$k_{\mathrm{in}}(4)=2$};
    \node[text=outorange, above=2pt, inner sep=0pt] at (6.65,7)
        {$k_{\mathrm{out}}(4)=4$};

    \foreach \x/\y in {1/6,2/2.5,3/3.5,5/1,6/3,7/2,8/5,9/7} {
        \draw[fill=barfill, draw=bardraw, line width=0.9pt]
            (\x-\w,0) rectangle (\x+\w,\y);
        \draw[axisgray, line width=0.7pt]
            (\x,0) -- (\x,-0.22)
            node[below=2pt, text=axisgray] {\small \x};
    }

    \draw[fill=focusfill, draw=focusdraw, line width=1.2pt]
        (4-\w,0) rectangle (4+\w,8);
    \draw[axisgray, line width=0.7pt]
        (4,0) -- (4,-0.22)
        node[below=2pt, text=focusdraw, font=\sffamily\bfseries] {\small 4};

    \draw[blockedgray, line width=0.85pt, densely dashed, ->]
        (4+\w,2) -- (7-\w,2);
    \node[text=blockedgray, font=\sffamily\bfseries, inner sep=0.5pt]
        at (6,2) {$\times$};
\end{tikzpicture}}
    \caption{
        Construction of the directed horizontal visibility graph.
        Horizontal visibility links are established between nodes $i$ and $j$ if $\min(r_i, r_j) > \max_{i < k < j} r_k$, and are oriented following the arrow of time.
        The blue incoming and orange outgoing arrows show the visible neighbors of node 4, giving $k_{\mathrm{in}}(4)=2$ and $k_{\mathrm{out}}(4)=4$.
        The dashed gray candidate link from node 4 to node 7 is blocked by node 6 because $r_6 \geq \min(r_4,r_7)$.
    }
    \label{fig:dhvg_example}
\end{figure}
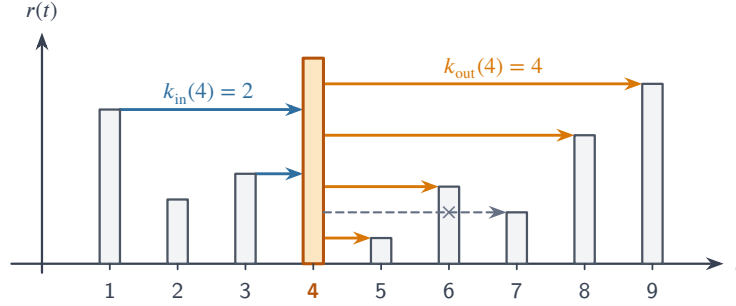

Figure~\ref{fig:dhvg_example} illustrates the HVG construction.
A directed edge was assigned from node $i$ to $j$ ($i < j$) if no intermediate observation blocked their horizontal line-of-sight:
\begin{equation}
    r_k < \min(r_i,r_j)
    \qquad \text{for all } i<k<j .
    \label{eq:hvg_condition}
\end{equation}
For a graph with $N$ nodes, the empirical in- and out-degree distributions are
\begin{equation}
    P_{\mathrm{in}}(k)
    = \frac{1}{N}\sum_{i=1}^{N}\mathbf{1}\!\left(k_i^{\mathrm{in}}=k\right),
    \qquad
    P_{\mathrm{out}}(k)
    = \frac{1}{N}\sum_{i=1}^{N}\mathbf{1}\!\left(k_i^{\mathrm{out}}=k\right),
    \label{eq:hvg_degree_distributions}
\end{equation}
where $k_i^{\mathrm{in}}$ and $k_i^{\mathrm{out}}$ are the in- and out-degrees of node $i$, respectively, and $\mathbf{1}(\cdot)$ is the indicator function.
For a time-reversible process, the in- and out-degree distributions are identical in the population. Finite-sample differences are therefore evaluated against surrogate null models.

The Kullback--Leibler divergence between two discrete distributions $P$ and $Q$ is defined as
\begin{equation}
    D_{\mathrm{KL}}(P\Vert Q)
    = \sum_k P(k)\ln\left[\frac{P(k)}{Q(k)}\right].
    \label{eq:hvg_kl}
\end{equation}
To capture the magnitude of this discrepancy, we employed the Jensen--Shannon divergence ($D_{\mathrm{JS}}$):
\begin{equation}
    D_{\mathrm{JS}}(P_{\mathrm{in}},P_{\mathrm{out}}) = \frac{1}{2}D_{\mathrm{KL}}(P_{\mathrm{in}}\Vert M) + \frac{1}{2}D_{\mathrm{KL}}(P_{\mathrm{out}}\Vert M),
    \label{eq:hvg_js}
\end{equation}
where $M(k)=\frac{1}{2}[P_{\mathrm{in}}(k)+P_{\mathrm{out}}(k)]$ is the mean degree distribution.
Conceptually, $D_\mathrm{JS}$ determines whether the return series exhibits disparate visibility statistics when viewed forward versus backward in time.
Thus, $D_{\mathrm{JS}}=0$ indicates identical in- and out-degree distributions, whereas larger values indicate a stronger in--out distributional mismatch.

However, because $D_\mathrm{JS}$ is symmetric, it does not identify which temporal side of the event drives the mismatch.
To retain this directional information, we also used the normalized signed KL imbalance ($A_\mathrm{KL}$),
\begin{equation}
    A_\mathrm{KL} =
    \frac{D_\mathrm{KL}(P_\mathrm{in}\,\|\,P_\mathrm{out}) - D_\mathrm{KL}(P_\mathrm{out}\,\|\,P_\mathrm{in})}
    {D_\mathrm{KL}(P_\mathrm{in}\,\|\,P_\mathrm{out}) + D_\mathrm{KL}(P_\mathrm{out}\,\|\,P_\mathrm{in})}.
    \label{eq:hvg_akl}
\end{equation}
Because empirical degree distributions can contain zero-probability bins, direct KL ratios may become unstable or undefined for degrees observed in only one direction.
Before computing the KL-based quantities, the in- and out-degree histograms were therefore regularized by adding a pseudocount of $0.5$ to each retained degree bin and then renormalizing the two histograms to probability distributions.
The normalization by the sum of both KL directions isolates the orientation of the mismatch from its magnitude.
A positive $A_\mathrm{KL}$ indicates that the in-degree distribution is less well represented by the out-degree distribution, whereas a negative value indicates the opposite.
Because edge direction follows time, this sign identifies whether the mismatch is weighted toward backward- or forward-visibility configurations and reverses under time reversal.
It does not, by itself, identify a unique buildup--recovery path.

Finally, we examined the degree-local asymmetry profile,
\begin{equation}
    A(k) = \frac{P_\mathrm{in}(k)-P_\mathrm{out}(k)}{P_\mathrm{in}(k)+P_\mathrm{out}(k)}.
    \label{eq:hvg_ak}
\end{equation}
$A(k)$ decomposes the signed asymmetry by degree.
At a given degree, $A(k)>0$ means that nodes with $k$ backward links occur more frequently than nodes with $k$ forward links. The opposite holds when $A(k)<0$.
This allows us to distinguish between imbalances arising from typical low-degree fluctuations and those concentrated in high-visibility return configurations.

The DHVG is constructed first from the signed, deseasonalized return series $\{r_t\}$ and then from $\{|r_t|\}$, $\{r_t:r_t>0\}$, and $\{-r_t:r_t<0\}$.
Absolute returns retain every observation after replacing each return by its magnitude, whereas the sign-conditioned subsequences preserve occurrence order but not elapsed clock time.
For every input, DHVG degree counts are computed separately within the same contiguous source-price timestamp segments and then pooled. Sign filtering does not reconnect events across a true source gap.
For the high-degree comparison, we restrict and renormalize the degree distributions at $k\geq10$.
This cutoff was chosen from the observed degree profiles. Results for $K=8$--$12$ and, for the KL-based measures, pseudocounts $0.1$ and $0.5$ are reported in the Supplementary Material.

\subsection{Surrogate-Based Decomposition}
\label{subsec:surrogate_decom}

To compare the observed diagnostics with constrained reference series, we applied shuffled and IAAFT surrogate analyses to the MF-DFA, CECP, and DHVG measures.
Shuffled surrogates preserve the marginal distribution of the original series while removing temporal dependence.
Iterative amplitude-adjusted Fourier transform (IAAFT) surrogates preserve both the marginal distribution and the linear autocorrelation structure while disrupting nonlinear phase relationships~\cite{Schreiber2000}.
Differences from shuffled surrogates reflect temporal ordering beyond the marginal distribution, and differences from IAAFT surrogates reflect structure beyond the linear autocorrelation.
These comparisons are defined relative to the constraints each surrogate retains and do not uniquely identify a generating mechanism.

For each asset and each surrogate type, we generated 100 realizations and recomputed the relevant diagnostic measures.
We assessed finite-realization effects by recomputing cumulative ensemble summaries and tail comparisons at 20, 50, and 100 realizations; the qualitative conclusions were unchanged between 50 and 100 realizations, so 100 realizations were retained for the reported comparisons.
The original estimates were compared with ensemble means, dispersions, and inclusive Monte Carlo tail probabilities, with mean $\pm$ one standard deviation used only descriptively.
For HVG asymmetry, this dispersion was computed at the level of the final diagnostic: for each surrogate realization $s$, we first evaluated $A_{\mathrm{KL}}^{(s)}$ from its own pair of KL divergences and then reported the mean and standard deviation of $\{A_{\mathrm{KL}}^{(s)}\}$.
The inclusive Monte Carlo comparisons and the corresponding multiplicity conventions are reported in Supplementary Tables~\ref*{tab:price_contiguous_k_grid} and~\ref*{tab:sampling_interval_combined}.

\section{Stylized Facts}
\label{sec:global}

\subsection{Distributional and Monofractal Properties}
\label{subsec:standard_facts}

We begin by characterizing the stylized facts across the cryptocurrencies and the S\&P~500 benchmark to evaluate the extent of their statistical overlap and identify where they leave room for structural diagnostics.
Accordingly, we conduct a heavy-tail analysis and estimate global Hurst exponents using second-order detrended fluctuation analysis (DFA-2).
In the heavy-tail analysis, we use the maximum likelihood estimation (MLE) based power-law fit following the Clauset--Shalizi--Newman procedure~\cite{Clauset_2009} to check whether all four assets occupy a common fat-tailed regime.

\begin{figure}[pos=h]
    \centering
    \includegraphics[width=\textwidth]{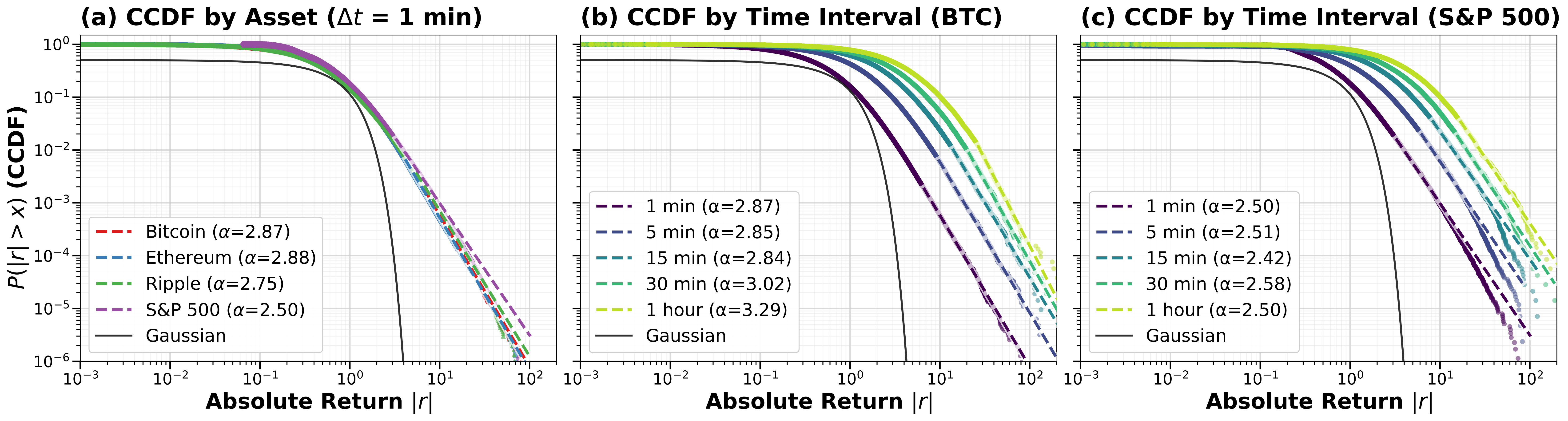}
    \caption{Complementary cumulative distribution functions (CCDFs) of absolute returns for cryptocurrencies and the S\&P~500 benchmark, with power-law fits.
        (a) CCDFs for one-minute, intraday-deseasonalized returns. The horizontal axis represents the absolute return threshold $x$ in these rescaled units, and the vertical axis represents the exceedance probability $P(|r| > x)$.
        Fat tails are observed for all assets, in clear contrast to the Gaussian distribution (black solid line).
        (b) Bitcoin and (c) S\&P~500 CCDFs across time aggregation intervals from $1$ to $60$ minutes.
        Colors change from purple to yellow as the time interval increases.
        The reported $\alpha$ values are cumulative Pareto tail-slope exponents estimated via maximum likelihood estimation (MLE).
    }
    \label{fig:ccdf}
\end{figure}

Figure~\ref{fig:ccdf} summarizes these tail characteristics, plotting the cumulative probability $P(|\tilde{r}| > x)$ against the return threshold $x$ on a double-logarithmic scale.
As shown in panel (a), the one-minute empirical complementary cumulative distribution functions (CCDFs) of all four assets decay much slower than a Gaussian (black solid line), confirming the presence of heavy-tailed returns.
The tail exponents estimated from the maximum likelihood estimation (MLE) fits fall in the range $\alpha \approx 2.5\text{--}2.9$.
In the econophysics literature, this is close to the inverse cubic law of return distributions ($\alpha \approx 3.0$), which describes the universal scale-invariant tail behavior observed across diverse financial markets~\cite{Mantegna1999, Cont2001}.
These selected assets exhibit broadly similar heavy-tailed behavior over the fitted range, departing from a Gaussian reference.

Panels (b) and (c) display the scale dependence of the tail behavior for Bitcoin and the S\&P~500, respectively.
As the aggregation interval increases from 1 to 60 minutes (indicated by the purple-to-yellow transition), both assets exhibit a rightward shift in their CCDFs.
For Bitcoin (panel b), the tail exponent exhibits a moderate increase from $\alpha \approx 2.8$ to $\alpha \approx 3.3$, indicating a gradual onset of aggregational Gaussianity.
In contrast, the S\&P~500 exponent (panel c) remains stable around $\alpha \approx 2.4\text{--}2.6$, showing no pronounced change within this short aggregation range.
Crucially, due to strong temporal dependence, this convergence toward the Gaussian attractor is extremely slow for both assets, keeping them well within the heavy-tailed regime ($\alpha < 4$) even at the hourly scale.
The tail estimates describe shared non-Gaussian features. Similar slopes do not establish equal risk, since scales, tail thresholds, and temporal dependence can differ.
Table~\ref*{tab:supp_tail_post2022} confirms that this heavy-tailed regime persists under a stricter post-2022 window.
The S\&P~500 slope shifts closer to the inverse-cubic benchmark in this window, underscoring the sensitivity of $\alpha$ to the sample choice.

We next use detrended fluctuation analysis (DFA) to test whether global monofractal scaling separates the asset classes.
The focus is on whether price directionality, volatility clustering, and trading activity exhibit common scaling patterns.
Table~\ref{tab:hurst} summarizes the log--log slopes of the fluctuation functions as global Hurst exponents ($H$) and compares their temporal origins using Shuffle and IAAFT surrogate tests.

\begin{table}[pos=htbp]
    \centering
    \caption{
        Global Hurst exponents ($H$) estimated by DFA-2 with second-order polynomial detrending.
        The table summarizes the Hurst exponents for returns ($r_t$), absolute returns ($|r_t|$), and trading volume ($V_t$) for Bitcoin (BTC), Ethereum (ETH), Ripple (XRP), and the S\&P~500 benchmark.
        Errors ($\pm$) denote the standard error (SE) of the log--log linear regression slope. The Shuffled and IAAFT errors are the SEs from linear regressions fitted to the 100-realization ensemble-mean fluctuation functions.
        Returns remain close to the random-walk $H \approx 0.5$, consistent with the absence of linear autocorrelation, whereas absolute returns and trading volume exhibit strong long-memory behavior with $H \gg 0.5$.
    }
    \label{tab:hurst}
    \setlength{\tabcolsep}{8pt}
    \renewcommand{\arraystretch}{0.9}
    \begin{tabular}{l l cccc}
        \toprule
        \textbf{Metric} & \textbf{Data Type}           & \textbf{BTC}        & \textbf{ETH}        & \textbf{XRP}        & \textbf{S\&P~500}   \\
        \midrule
        \multirow{3}{*}{Returns}
                        & Original ($H_{\text{orig}}$) & $0.4964 \pm 0.0020$ & $0.5033 \pm 0.0020$ & $0.4996 \pm 0.0025$ & $0.4923 \pm 0.0022$ \\
                        & Shuffled ($H_{\text{shuf}}$) & $0.5064 \pm 0.0016$ & $0.5069 \pm 0.0016$ & $0.5067 \pm 0.0016$ & $0.5067 \pm 0.0017$ \\
                        & IAAFT ($H_{\text{IAAFT}}$)   & $0.4958 \pm 0.0019$ & $0.5021 \pm 0.0018$ & $0.4966 \pm 0.0022$ & $0.4911 \pm 0.0020$ \\
        \midrule
        \multirow{3}{*}{Absolute Returns}
                        & Original ($H_{\text{orig}}$) & $0.9010 \pm 0.0057$ & $0.8999 \pm 0.0057$ & $0.9079 \pm 0.0052$ & $0.9215 \pm 0.0108$ \\
                        & Shuffled ($H_{\text{shuf}}$) & $0.5068 \pm 0.0016$ & $0.5065 \pm 0.0016$ & $0.5068 \pm 0.0016$ & $0.5073 \pm 0.0017$ \\
                        & IAAFT ($H_{\text{IAAFT}}$)   & $0.8966 \pm 0.0055$ & $0.8971 \pm 0.0055$ & $0.9058 \pm 0.0051$ & $0.9124 \pm 0.0099$ \\
        \midrule
        \multirow{3}{*}{Volume}
                        & Original ($H_{\text{orig}}$) & $0.8839 \pm 0.0020$ & $0.8893 \pm 0.0019$ & $0.9315 \pm 0.0028$ & $0.8101 \pm 0.0119$ \\
                        & Shuffled ($H_{\text{shuf}}$) & $0.5063 \pm 0.0016$ & $0.5064 \pm 0.0016$ & $0.5068 \pm 0.0016$ & $0.5072 \pm 0.0017$ \\
                        & IAAFT ($H_{\text{IAAFT}}$)   & $0.8751 \pm 0.0020$ & $0.8831 \pm 0.0015$ & $0.9287 \pm 0.0020$ & $0.8205 \pm 0.0098$ \\
        \bottomrule
    \end{tabular}
\end{table}

Table~\ref{tab:hurst} summarizes the DFA-2 scaling exponents ($H$) for returns ($r_t$), absolute returns ($|r_t|$), and trading volume ($V_t$) against their shuffled and IAAFT surrogate counterparts.
For the return series, the original exponents ($H_{\text{orig}}$) and both surrogate results ($H_{\text{shuf}}$, $H_{\text{IAAFT}}$) all cluster tightly around the theoretical random-walk benchmark ($H = 0.5$), hovering in the narrow range of $0.49\text{--}0.51$.
These estimates are close to the uncorrelated scaling benchmark. A DFA exponent near $0.5$ alone does not establish independence, directional unpredictability, or weak-form market efficiency.

By contrast, absolute returns and trading volume both exhibit strong long-range persistence across all assets.
Specifically, the original Hurst exponents are consistently high for both absolute returns ($H_{\text{orig}} \approx 0.90\text{--}0.92$) and trading volume ($H_{\text{orig}} \approx 0.81\text{--}0.93$), demonstrating persistent volatility clustering and trading activity.
For both metrics, this long-range correlation is entirely destroyed by shuffling ($H_{\text{shuf}} \approx 0.51$) but preserved by the IAAFT surrogates ($H_{\text{IAAFT}} \approx H_{\text{orig}}$).
The IAAFT comparisons indicate that the retained power spectra reproduce much of the observed DFA persistence in these observables, providing a baseline for their broad scaling similarity.
Replicate-level comparisons likewise do not support a general crypto--equity separation (Supplementary Material).

\subsection{Multifractal Analysis}
\label{subsec:mfdfa}

\begin{figure}[pos=htbp]
    \centering
    \includegraphics[width=\textwidth]{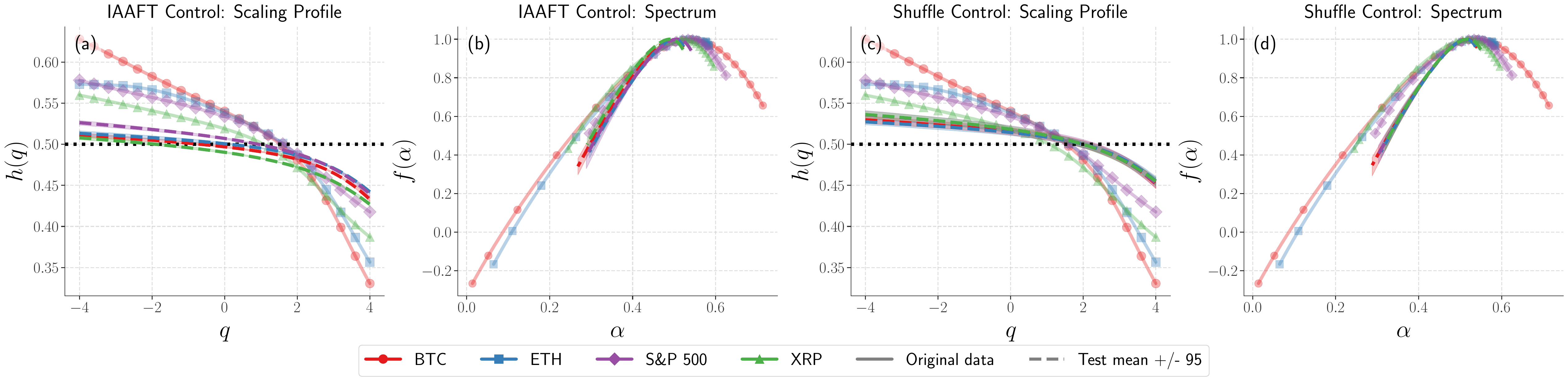}
    \caption{
        MF-DFA surrogate comparison for the return series.
        Panels (a)--(b) show the IAAFT comparison for the generalized Hurst exponent $h(q)$ and the multifractal spectrum $f(\alpha)$, respectively.
        Panels (c)--(d) show the corresponding Shuffle comparison.
        Solid curves denote the original series. Dashed curves and shaded bands denote surrogate ensemble means and uncertainty bands.
    The close agreement between the Shuffle and IAAFT baselines indicates that the surrogate multifractal profile is dominated by the marginal distribution, while the remaining original--IAAFT displacement is treated as an operational residual for temporal organization.
    }
    \label{fig:mfdfa}
\end{figure}

These single-scale estimates summarize broadly similar scaling but do not establish equivalent temporal organization.
We therefore apply multifractal detrended fluctuation analysis (MF-DFA) with Shuffle and IAAFT surrogate tests to examine this scale-dependent heterogeneity.
By comparing the empirical scaling of the original series against their shuffled and IAAFT surrogate ensembles, this approach provides an operational comparison against marginal-distribution and linear-correlation references.
These comparison results are presented in Figure~\ref{fig:mfdfa} and Table~\ref{tab:multifractal}.

\begin{table}[pos=hb]
    \centering
    \caption{Multifractal parameters and surrogate comparison results by asset.
        The spectrum width $\Delta\alpha$ is computed over the moment range $q \in [-4, 4]$.
        The generalized Hurst exponents $h(q)$ emphasize small-amplitude fluctuations ($q=-3$), the logarithmic-average case ($q=0$), and large-amplitude fluctuations ($q=3$), comparing empirical returns against Shuffle and IAAFT surrogate ensemble means.
    }
    \label{tab:multifractal}
    \setlength{\tabcolsep}{3pt}
    \renewcommand{\arraystretch}{0.9}
    \resizebox{\columnwidth}{!}{%
        \begin{tabular}{l ccc ccc ccc ccc}
            \toprule
                     & \multicolumn{3}{c}{$\Delta\alpha$} & \multicolumn{3}{c}{$h(-3)$} & \multicolumn{3}{c}{$h(0)$} & \multicolumn{3}{c}{$h(3)$}                                                                                \\
            \cmidrule(lr){2-4}\cmidrule(lr){5-7}\cmidrule(lr){8-10}\cmidrule(lr){11-13}
            Asset    & Original                           & Shuffle                     & IAAFT                      & Original                   & Shuffle & IAAFT  & Original & Shuffle & IAAFT  & Original & Shuffle & IAAFT  \\
            \midrule
            BTC      & 0.7160                             & 0.2515                      & 0.2506                     & 0.6090                     & 0.5269  & 0.5080 & 0.5408   & 0.5152  & 0.4969 & 0.4164   & 0.4830  & 0.4654 \\
            ETH      & 0.5217                             & 0.2289                      & 0.2264                     & 0.5746                     & 0.5249  & 0.5109 & 0.5384   & 0.5145  & 0.5009 & 0.4319   & 0.4842  & 0.4712 \\
            XRP      & 0.3502                             & 0.2309                      & 0.2322                     & 0.5519                     & 0.5327  & 0.5049 & 0.5197   & 0.5184  & 0.4909 & 0.4297   & 0.4825  & 0.4553 \\
            S\&P~500 & 0.3181                             & 0.2394                      & 0.2475                     & 0.5692                     & 0.5314  & 0.5224 & 0.5340   & 0.5173  & 0.5070 & 0.4549   & 0.4834  & 0.4714 \\
            \bottomrule
        \end{tabular}%
    }
\end{table}

Figure~\ref{fig:mfdfa} compares the MF-DFA results of the original return series against their IAAFT and Shuffle surrogates across two representations: the generalized Hurst exponent $h(q)$ and the multifractal spectrum $f(\alpha)$.
For the original data, the decreasing $h(q)$ and broad, concave $f(\alpha)$ spectra indicate heterogeneous scaling over the fitted range.
In contrast, both surrogate ensembles produce flatter $h(q)$ curves and narrower spectra under the reported settings.
The close overlap between the Shuffle and IAAFT curves indicates that the surrogates' remaining scaling behavior is primarily driven by the marginal distribution of returns, while the displacement between the original and IAAFT curves is treated as an operational residual for temporal organization.

These scaling properties are quantitatively summarized in Table~\ref{tab:multifractal}.
For all assets, negative moments yield exponents above $0.5$, whereas positive moments yield exponents below $0.5$. Negative and positive moments emphasize small- and large-amplitude fluctuations, respectively. They do not designate separate short- and long-time scaling ranges.
Bitcoin exhibits the strongest $q$-dependence and the widest spectrum ($\Delta\alpha = 0.7160$), followed by Ethereum ($0.5217$), while XRP ($0.3502$) and the S\&P~500 ($0.3181$) show much narrower spectra.
By comparison, the Shuffle and IAAFT surrogates hover near the mono-scaling benchmark ($h(q) \approx 0.5$) and yield remarkably consistent spectrum widths ($\Delta\alpha \approx 0.23$--$0.25$) across all assets.
The wider original spectra indicate scaling structure not reproduced by these controls, but finite-size effects, heavy-tailed marginals, nonstationarity, volatility-regime variation, and the fitting range can all affect multifractal estimates.
We therefore do not interpret a wider spectrum as, by itself, confirmation of genuine temporal dynamics or as a unique separation of nonlinear and heavy-tail effects.

To summarize these comparisons, we write the empirical generalized Hurst exponent as
\begin{equation}
    h_\mathrm{orig}(q) = h_\mathrm{shuf}(q) + \Delta_\mathrm{lin}(q) + \Delta_\mathrm{nl}(q),
    \label{eq:hq_decomp}
\end{equation}
where $h_\mathrm{shuf}(q)$ is the shuffled reference, $\Delta_\mathrm{lin}(q)=h_\mathrm{IAAFT}(q)-h_\mathrm{shuf}(q)$ is the difference between the two surrogate references, and $\Delta_\mathrm{nl}(q)=h_\mathrm{orig}(q)-h_\mathrm{IAAFT}(q)$ is the original--IAAFT residual. These labels denote an operational comparison, not a unique causal decomposition.

The original--IAAFT width residual is positive for all assets but differs substantially in magnitude.
It reaches $0.4654$ for BTC and $0.2953$ for ETH, compared with $0.1180$ for XRP and $0.0706$ for the S\&P~500.
Likewise, the empirical multifractal width varies more strongly across individual assets than across the simple cryptocurrency--equity distinction, with XRP exhibiting scaling characteristics closer to the S\&P~500 than to BTC.
These results indicate asset-specific scaling differences but not a clear asset-class separation.
Figure~\ref*{fig:supp_mfdfa} confirms that BTC retains the broadest spectrum under a stricter post-2022 window, though the relative ordering of ETH, XRP, and S\&P~500 is not preserved as ETH's spectrum narrows the most.
The shuffle and IAAFT surrogates remain in close agreement in this window as well, consistent with MF-DFA's use as a sensitivity diagnostic rather than a crypto--equity discriminator.
This observation motivates the subsequent ordinal-pattern and visibility-network analyses, which examine temporal organization beyond conventional multifractal scaling.

\section{Structural and Network Diagnostics}
\label{sec:structural_diagnostics}

While the four assets share common stylized facts---including heavy-tailed distributions, volatility clustering, and similar multifractal scaling properties---the original--IAAFT differences in MF-DFA provide a descriptive contrast with the surrogate controls, but do not by themselves establish residual temporal dynamics beyond linear correlations.
We therefore employ the complexity--entropy causality plane (CECP) and the directed horizontal visibility graph (DHVG) to probe structural differences that remain hidden from these global indicators.
We compare each diagnostic with shuffled surrogates, which remove all temporal dependence, and IAAFT surrogates, which preserve linear autocorrelation, thereby isolating overall temporal structure relative to Shuffle and nonlinear structure relative to IAAFT.
This comparison does not, however, attribute the residual to a specific nonstationary or nonlinear generating mechanism.

\subsection{Ordinal Pattern Structure}
\label{subsec:cecp_results}

We examine short-horizon ordinal structure in the $(1-H,C)$ plane, where proximity to the origin indicates higher entropy and weaker symbolic organization.
Results are reported for $D=5$ and $\tau=1$.

\begin{figure}[pos=htbp]
    \centering
    \includegraphics[width=0.9\textwidth]{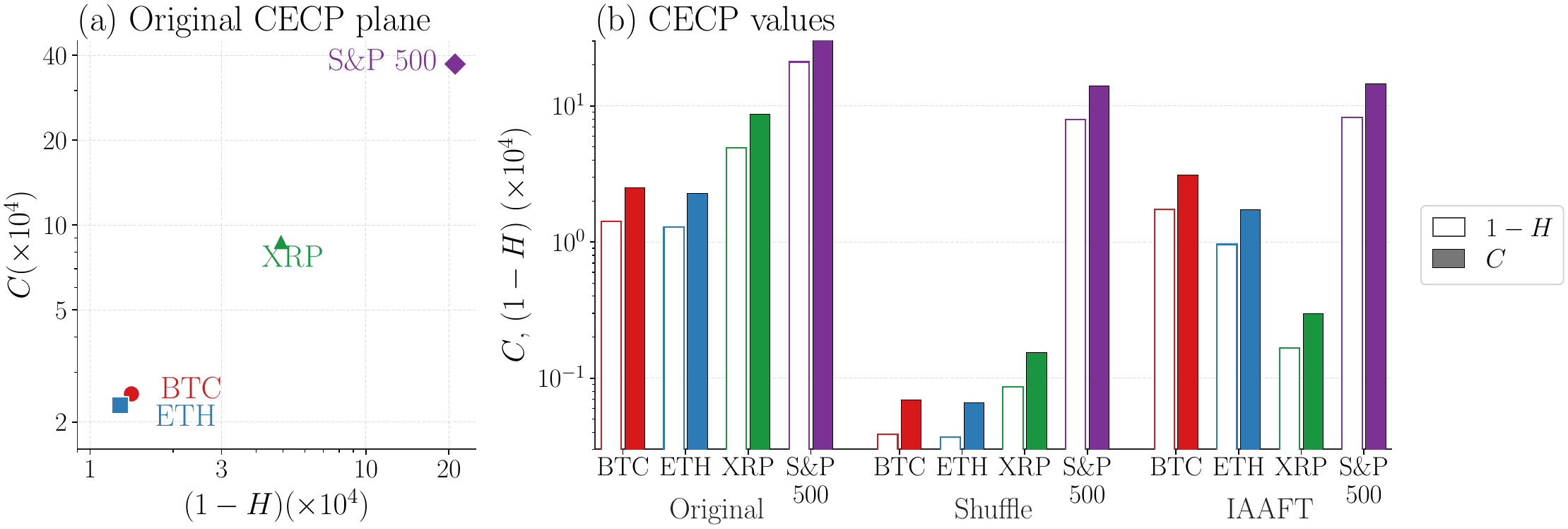}
    \caption{
        Complexity--entropy causality plane (CECP) analysis of asset returns.
        (a) Locations of BTC, ETH, XRP, and the S\&P~500 in the $(1-H,C)$ plane.
        (b) Values of $1-H$ (hollow bars) and $C$ (filled bars) for the original, shuffled, and IAAFT surrogate series.
        All values are scaled by $10^4$.
        Shuffling destroys temporal ordering while preserving the empirical marginal distribution, whereas IAAFT surrogates also approximately preserve the linear power spectrum.
        The cryptocurrencies lie closer to the random limit than the S\&P~500. The surrogate comparisons show asset-dependent original--surrogate gaps.
    }
    \label{fig:cecp_2020}
\end{figure}

Figure~\ref{fig:cecp_2020}(a) shows the original asset returns in the $(1-H, C)$ plane.
The permutation entropy $H$ measures the diversity and evenness of ordinal patterns, reaching its maximum when all possible patterns occur equally often.
The statistical complexity $C$ is low when the pattern distribution is either uniform or dominated by a single pattern, and becomes larger when multiple patterns occur with unequal frequencies.
BTC, ETH, and XRP cluster near the high-entropy, low-complexity boundary ($H \to 1, C \to 0$), indicating greater local ordinal randomness.
Although XRP is slightly displaced from BTC and ETH, the three cryptocurrencies remain clearly separated from the S\&P~500, which has a larger entropy deficit and statistical complexity.
In Figure~\ref*{fig:supp_cecp_grid}, the asset ordering underlying this separation remains stable across a broader $D\in\{3,\dots,7\}$, $\tau\in\{1,2,3,5,10\}$ embedding grid.

The surrogate analysis in Figure~\ref{fig:cecp_2020}(b) clarifies the statistical origins of this separation by comparing $1-H$ and $C$ across original, shuffled, and IAAFT surrogate ensembles.
Under shuffling, the cryptocurrency coordinates approach the random limit, whereas the S\&P~500 retains a non-zero baseline.
This residual baseline is consistent with the higher frequency of zero-return intervals in the S\&P~500 series, which is preserved by shuffling and produces repeated ordinal ties.

The IAAFT comparison assesses how much of the ordinal displacement is reproduced when the linear spectrum is approximately retained.
For BTC and ETH, the IAAFT surrogates reproduce most of the original complexity and entropy deficit, indicating a greater contribution from linear correlations than for XRP.
For XRP, the IAAFT coordinates lie closer to the shuffled baseline than those of BTC and ETH, indicating a smaller contribution from linear correlations.
Its correspondingly larger original--IAAFT gap indicates ordinal structure not reproduced by that surrogate constraint.
For the S\&P~500, the Shuffle and IAAFT baselines are nearly identical, while a substantial original--IAAFT gap remains.
In Figure~\ref*{fig:supp_cecp}, the cross-asset CECP hierarchy and its surrogate-based decomposition are both preserved under a stricter post-2022 window.
Sampling-interval sensitivity of this CECP comparison is evaluated over the same 1-, 5-, and 10-minute range and reported in Supplementary Sec.~\ref*{subsec:supp_cecp_sampling}: the asset-dependent original--IAAFT gap seen here at one minute is not preserved at coarser intervals.

\subsection{Directed Horizontal Visibility Graph (DHVG)}
\label{subsec:hvg_results}

Having established differences in local ordinal structure, we use the Directed Horizontal Visibility Graph (DHVG) to examine how time irreversibility varies with node degree.
Unlike the CECP, which characterizes local symbolic ordering, the DHVG maps relative visibility and identifies the high-degree observations that dominate temporal asymmetry.

\begin{figure}[pos=htbp]
    \centering
    \includegraphics[width=\textwidth]{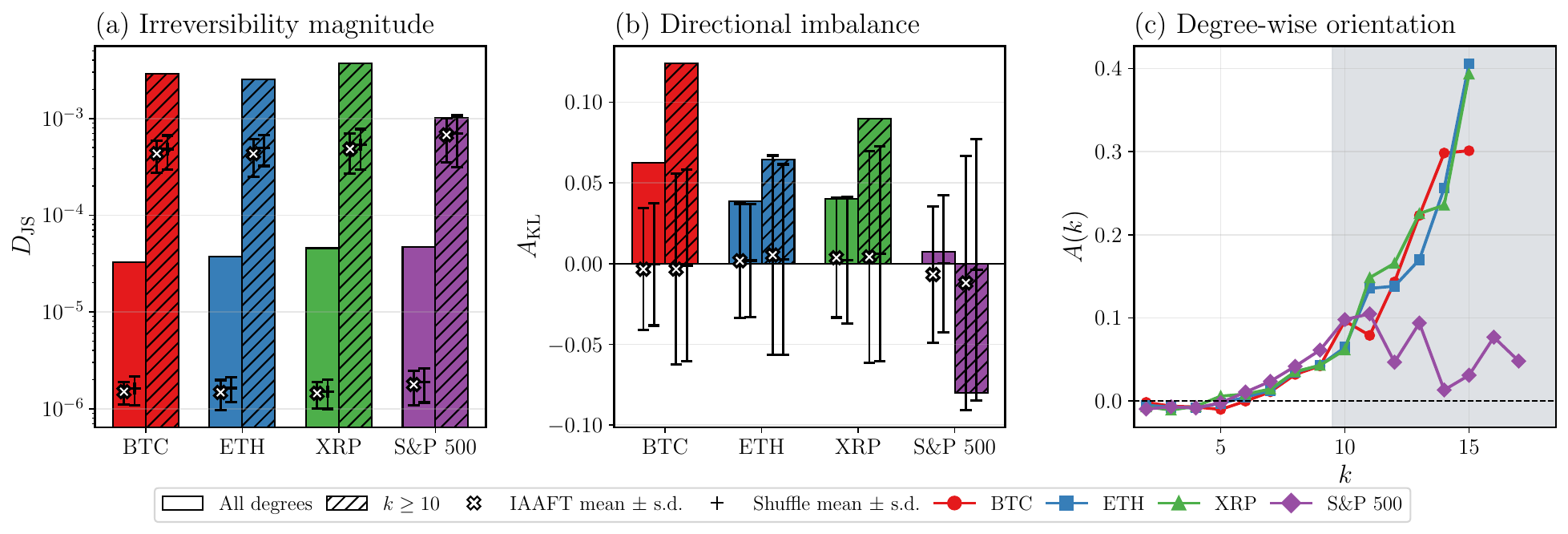}
    \caption{
        Directed-HVG decomposition of time irreversibility and directional asymmetry with source-timestamp boundaries preserved.
        Panels show (a) Jensen--Shannon divergence $D_\mathrm{JS}$ between the in- and out-degree distributions, measuring total time irreversibility; (b) signed directional KL asymmetry $A_\mathrm{KL}$; and (c) degree-wise asymmetry $A(k)$, which localizes directional imbalance by node degree.
        In panels (a) and (b), solid and hatched bars denote the observed values over all degrees and over the high-degree sector ($k\geq10$), respectively.
        Open crosses and plus markers indicate the IAAFT and shuffled surrogate means, with error bars showing one standard deviation.
        In panel (c) the gray region marks $k\geq10$, and only well-supported degrees ($n_k=c_\mathrm{in}(k)+c_\mathrm{out}(k)\geq100$) are shown as filled markers connected by solid lines.
    }
    \label{fig:hvg_decomposition}
\end{figure}

In Figure~\ref{fig:hvg_decomposition}(a), the observed $D_\mathrm{JS}$ exceeds both surrogate means for every asset over all degrees, an aggregate discrepancy not reproduced by the controls.
The nearly overlapping shuffled and IAAFT baselines indicate little additional contribution from linear temporal correlations.
At this aggregate level, however, $D_\mathrm{JS}$ does not clearly separate cryptocurrencies from the S\&P~500.

Figure~\ref{fig:hvg_decomposition}(c) shows that $A(k)$ begins to depart from zero around $k=6$--10 for all four assets, an onset that is not specific to cryptocurrencies.
Beyond this transition, however, the cryptocurrency profiles continue to increase into the well-supported high-degree tail, whereas the S\&P~500 profile largely levels off, so the persistence of the asymmetry is asset-specific.
A more detailed degree-wise comparison in the Supplementary Material shows this persistence becoming statistically distinguishable from $k=12$ for Bitcoin and XRP and from $k=14$ for Ethereum.

\begin{figure}[pos=htbp]
    \centering
    \includegraphics[width=\textwidth]{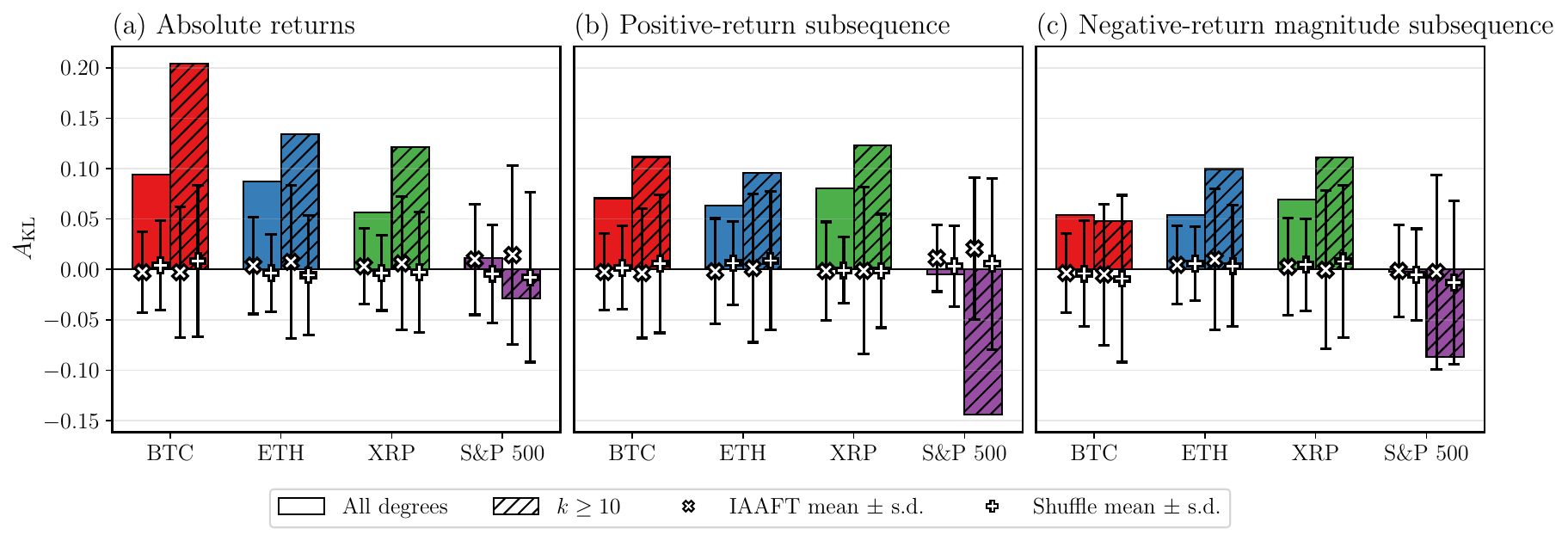}
    \caption{
        Directional HVG asymmetry for transformed returns.
        Panels show $A_\mathrm{KL}$ for (a) absolute returns, (b) positive-return subsequences, and (c) negative-return-magnitude subsequences.
        Solid and hatched bars denote observed values over all degrees and over $k\geq10$, respectively.
        Open crosses and plus markers denote IAAFT and shuffled surrogate means, with error bars showing one standard deviation.
        Absolute returns preserve all observations in the preprocessed input order after removing return signs.
        Sign-conditioned subsequences retain the order of same-sign returns after removing opposite-sign observations, but not the elapsed time between retained observations.
        For every transform, graphs are constructed separately within contiguous source-price segments and only their degree counts are pooled.
    }
    \label{fig:hvg_transform_akl}
\end{figure}

Restricting the analysis to this high-degree sector, located via $A(k)$, sharpens both diagnostics in Figure~\ref{fig:hvg_decomposition}(a,b): $D_\mathrm{JS}$ becomes nominally significant for each cryptocurrency under both null models, while the S\&P~500 value remains within the surrogate range, and $A_\mathrm{KL}$ shifts substantially further positive for the cryptocurrencies and negative for the S\&P~500.
That $A(k)$ locates a sector where $D_\mathrm{JS}$ turns significant and $A_\mathrm{KL}$ shifts this sharply makes the degree-resolved profile a meaningful diagnostic in its own right, independent of $A_\mathrm{KL}$'s own significance.
This separation persists under a stricter post-2022 window and 5-minute aggregation, but becomes indistinguishable at 10 minutes, consistent with a high-frequency origin for the asymmetry.
In DHVG terms, the positive high-degree $A_\mathrm{KL}$ indicates greater visibility toward the past than toward the future.
The raw-return analysis combines positive and negative moves, so it cannot determine whether the observed visibility asymmetry is associated primarily with magnitude ordering or with a particular return sign.
Figure~\ref{fig:hvg_transform_akl} separates these contributions using absolute returns, positive-return subsequences, and negative-return-magnitude subsequences.

In Figure~\ref{fig:hvg_transform_akl}(a), absolute-return $A_\mathrm{KL}$ exceeds the surrogate ranges for all three cryptocurrencies and increases when the analysis is restricted to $k\geq10$, most clearly for BTC, whereas the S\&P~500 remains within its surrogate range and becomes slightly negative.
This directional pattern indicates that return-magnitude ordering contributes to the high-degree asymmetry.

Figure~\ref{fig:hvg_transform_akl}(b) shows the same directional pattern for the positive-return subsequences: all three cryptocurrencies remain positive, with $A_\mathrm{KL}$ shifting further positive at $k\geq10$, and XRP moving outside its surrogate range. BTC and ETH stay within their ranges on the positive side, and the S\&P~500 stays within its own range on the negative side.
Because opposite-sign observations and elapsed time are removed, it does not imply an uninterrupted rally in clock time.

For negative-return magnitudes, Figure~\ref{fig:hvg_transform_akl}(c) shows positive all-degree $A_\mathrm{KL}$ for all three cryptocurrencies and a value close to zero for the S\&P~500.
At $k\geq10$, Ethereum and XRP remain more positive than Bitcoin, while the S\&P~500 becomes negative. All four values stay within their surrogate ranges at this degree level.
Because opposite-sign observations and elapsed time are removed, this pattern does not imply an uninterrupted sell-off in clock time.
Within the retained negative-return ordering, the relative high-degree orientation is therefore asset-dependent.
This consistency across transforms points to magnitude ordering, rather than return sign, as the likely contributor to the high-degree asymmetry, though no transform separates all three cryptocurrencies uniformly. Degree-resolved results for each transformation are reported in the Supplementary Material.

\section{Discussion and Conclusion}
\label{sec:conclusion}

This study examined whether the macroscopic similarities between cryptocurrencies and the S\&P~500 extend to their temporal organization.
Distributional, DFA, and MF-DFA analyses showed that all four assets share heavy-tailed returns, volatility clustering, and multifractal scaling.
The strength of multifractality, however, varied more across individual assets than between cryptocurrencies and the equity benchmark.
Since these measures did not clearly separate the two asset classes, we next examined short-horizon temporal structure using surrogate-controlled CECP and DHVG analyses.
The CECP revealed weaker local ordinal organization in cryptocurrencies than in the S\&P~500, whereas the DHVG localized the crypto--equity separation to high-visibility events.
Conventional stylized facts therefore coexist with diagnostic-specific differences in local ordinal organization and event-centered time asymmetry.

The CECP surrogate comparisons show that the sources of local ordinal organization differ across assets.
This heterogeneity may reflect distinct trading environments.
Standardized execution and liquidity provision can generate recurrent ordering in equity-index returns, whereas fragmented and continuously operating cryptocurrency markets may weaken such regularities.

Under the primary one-minute, intraday-rescaled specification, weaker local ordinal organization in cryptocurrencies coexists with a larger high-degree visibility mismatch relative to matched controls.
This mismatch is a directional visibility statistic. It does not by itself identify quiet pre-event paths, abrupt onsets, or persistence after an event.
The $D_\mathrm{JS}$ pattern after sign removal indicates that return-magnitude ordering contributes to the mismatch beyond the marginal distribution and linear spectrum retained by the IAAFT surrogates (Supplementary Material).
At the descriptive level, sign conditioning shows a shared upside but an asset-dependent downside component.

Accordingly, models that reproduce tails, volatility clustering, and scaling can still miss the event-ordering structure captured by the CECP and DHVG diagnostics.
Model adequacy should therefore be assessed against the CECP coordinates and DHVG signatures $D_\mathrm{JS}$, $A_\mathrm{KL}$, and $A(k)$, rather than conventional stylized facts alone.

This scope is limited to three cryptocurrencies and one equity-index futures benchmark during 2020--2025.
Over the same sampling-interval range, both the CECP original--IAAFT displacement and the high-degree $D_\mathrm{JS}$ separation lose their asset-dependent contrast as the interval coarsens from one to 10 minutes (Supplementary Sec.~\ref*{subsec:supp_cecp_sampling}).
This shared weakening points to a high-frequency origin for the asset-dependent contrast, rather than to a sampling-independent market characteristic.

\bibliographystyle{elsarticle-num-names}
\bibliography{references}

\end{document}


\let\WriteBookmarks\relax
\shorttitle{Supplementary Material: Universality and Heterogeneity of Stylized Facts}
\shortauthors{Kim et al.}

\title [mode = title]{Supplementary Material: Universality and Heterogeneity of Stylized Facts in Cryptocurrency and Equity Markets}

\author[1]{Jaesung Kim}[orcid=0009-0008-7268-8576]

\author[1]{Changhee Cho}

\author[1,2,3]{Jae Woo Lee}
\cormark[1]
\ead{jaewlee@inha.ac.kr}

\cortext[1]{Corresponding author}

\affiliation[1]{organization={Department of Physics, Inha University},
    city={Incheon},
    postcode={22212},
    country={Republic of Korea}}

\affiliation[2]{organization={Institute of Quantum Science, Inha University},
    city={Incheon},
    postcode={22212},
    country={Republic of Korea}}

\affiliation[3]{organization={School of Computational Sciences, Korea Institute for Advanced Study},
    city={Seoul},
    postcode={02455},
    country={Republic of Korea}}

\begin{abstract}
    This supplementary material provides the technical foundations and robustness checks for the main analysis.
    It details the mathematical formulations and algorithms for the CECP, directed HVG, and surrogate data generation.
    It also documents data construction and robustness tests, including the post-2020 cutoff justification, post-2022 re-analyses, CECP grid search, degree-resolved HVG decomposition, degree-threshold and pseudocount sensitivity, and sampling-interval sensitivity.
\end{abstract}

\maketitle

\section{Supplementary Overview}
\label{sec:supp_overview}

This supplementary material contains the technical foundations and robustness checks for the main analysis.
We summarize data construction in Sec.~\ref{sec:supp_data_construction}, detail the mathematical formulations for the CECP and directed HVG in Secs.~\ref{sec:supp_cecp} and~\ref{sec:supp_hvg}, and describe surrogate data generation in Sec.~\ref{sec:supp_surrogates}.
We also present robustness tests regarding the post-2020 sample cutoff, the post-2022 re-analysis, the CECP embedding parameters, the HVG degree threshold and pseudocount, and the sampling interval.

Specifically, we justify the January~1, 2020 maturity cutoff in Sec.~\ref{sec:supp_scope} using rolling DFA, volatility memory, and GARCH parameters.
We repeat the tail, MF-DFA, CECP, and HVG diagnostics for the post-2022 period in Sec.~\ref{sec:supp_post2022}.
We also test the sensitivity of the CECP to embedding parameters in Sec.~\ref{sec:supp_cecp_grid}, and examine HVG irreversibility by degree, return transform, threshold, and pseudocount within Sec.~\ref{sec:supp_hvg_degree}.
Sec.~\ref{sec:revision_sampling} then tests both the DHVG and CECP diagnostics across the 1-, 5-, and 10-minute sampling intervals, the temporal aggregation axis under which the crypto--equity separation is least stable.
These checks assess whether the descriptive ordinal and topological contrasts of the selected assets persist under the tested windows, embeddings, and sampling rates.

\section{Data Construction}
\label{sec:supp_data_construction}

Cryptocurrency weekends are retained where source prices are available, whereas the futures data follow their exchange trading calendar.
Non-finite observations and non-trading intervals in this primary series are not filled or bridged.
The continuous futures series uses volume-based contract selection and ratio backward adjustment at rollovers.
Trading volume is analyzed separately from the return series.

For the primary signed DHVG analysis (and, identically, the DFA-2, MF-DFA, and CECP robustness recomputations below), returns are reconstructed from the earliest available source-price records under a single shared gap rule: a missing 1-minute tick is filled at the last observed price (return exactly 0) when the surrounding source-price gap is 2--14 minutes, and left unfilled---starting a new contiguous segment, within which alone patterns/windows/graph edges are ever formed---when the gap is 15 minutes or longer.
This threshold follows directly from the gap-length audit below: gaps of 2--14 minutes occur at essentially every clock time with no recurring structure (incidental thin-liquidity or feed misses), whereas gaps of 15 minutes or longer are either a recognizable session boundary (the S\&P~500's daily overnight halt, weekend, and holiday closures) or, for its recurring $\sim$15--16 minute pre-close pause clustered at exactly 16:14 America/New\_York, consistent with a settlement/quote-freeze window distinct from continuous trading rather than a data gap.
Table~\ref{tab:data_desc} reports, for each asset, the per-exchange start date and common end date of the analyzed sample together with the resulting observation and segment counts under this rule.

For the DHVG analyses, each return transformation is constructed separately within each contiguous source-price segment.
The resulting in- and out-degree counts are then pooled across segments for the reported degree distributions and diagnostics; no graph edge spans a source-price gap.

\begin{table}[pos=htbp]
    \centering
    \caption{
        Per-exchange data coverage, and sample sizes retained for the primary signed one-minute source-price reconstruction shared by DFA-2, MF-DFA, CECP, and DHVG.
        A segment is a maximal run within which no source-price gap of 15 minutes or longer occurs.
        Shorter gaps (2--14 minutes) are filled at zero return and do not start a new segment.
        The post-2020 production window used throughout the main text begins after all three cryptocurrency assets are simultaneously covered across this exchange universe (Sec.~\ref{sec:supp_scope}).
    }
    \label{tab:data_desc}
    \setlength{\tabcolsep}{6pt}
    \begin{tabular}{l l c c c c}
        \toprule
         & & \multicolumn{2}{c}{Exchange coverage} & \multicolumn{2}{c}{Shared source-price input} \\
        \cmidrule{3-4} \cmidrule{5-6}
        Asset & Exchange & Start Date & End Date & Observations & Segments \\
        \midrule
        \multirow{3}{*}{BTC} & Bitstamp & Jan 1, 2015 & \multirow{3}{*}{Dec 31, 2025} & \multirow{3}{*}{3,156,479} & \multirow{3}{*}{1} \\
         & Coinbase & Jan 1, 2015 & & & \\
         & Kraken   & Jan 1, 2015 & & & \\
        \midrule
        \multirow{3}{*}{ETH} & Bitstamp & Sep 1, 2017 & \multirow{3}{*}{Dec 31, 2025} & \multirow{3}{*}{3,156,479} & \multirow{3}{*}{1} \\
         & Coinbase & Jun 1, 2016 & & & \\
         & Kraken   & Jan 1, 2016 & & & \\
        \midrule
        \multirow{3}{*}{XRP} & Bitstamp & Jan 1, 2017 & \multirow{3}{*}{Dec 31, 2025} & \multirow{3}{*}{3,156,220} & \multirow{3}{*}{12} \\
         & Coinbase & Mar 1, 2019 & & & \\
         & Kraken   & Jun 1, 2017 & & & \\
        \midrule
        S\&P~500 & CME & Jan 1, 2015 & Dec 31, 2025 & 2,117,623 & 1,924 \\
        \bottomrule
    \end{tabular}
\end{table}

\section{Complexity--Entropy Causality Plane}
\label{sec:supp_cecp}

\subsection{Phase Space Reconstruction: The Bandt--Pompe Mapping}
To analyze the temporal organization of a time series $\mathcal{X} = \{x_t : t = 1, \dots, N\}$, we first transform the continuous-valued data into a sequence of discrete symbolic patterns.
This process requires two parameters: the embedding dimension $D$ (pattern length) and the time delay $\tau$ (sampling interval).

For each time index $s$, we define a $D$-dimensional vector:
\begin{equation}
    X_s^{(D, \tau)} = \left( x_s, x_{s+\tau}, \dots, x_{s+(D-1)\tau} \right).
\end{equation}
The core idea of the Bandt--Pompe method~\cite{Bandt2002} is to consider the relative order of the elements within this vector.
Each vector is mapped to an ordinal pattern (permutation) $\pi = (r_0, r_1, \dots, r_{D-1})$ such that the elements of $X_s^{(D, \tau)}$ are ordered as:
\begin{equation}
    x_{s+r_0\tau} \leq x_{s+r_1\tau} \leq \dots \leq x_{s+r_{D-1}\tau}.
\end{equation}
In the rare case of perfectly equal values ($x_{s+i\tau} = x_{s+j\tau}$), we maintain their chronological order ($i < j$).

\paragraph{Example:} For $D=3, \tau=1$, a segment $(x_s, x_{s+1}, x_{s+2}) = (15, 7, 22)$ satisfies $x_{s+1} < x_s < x_{s+2}$.
This corresponds to the permutation $(1, 0, 2)$ among the $3! = 6$ possible states.

\subsection{Probability Distribution of Ordinal Patterns}
By sliding the $D$-length window across the entire time series, we count the occurrences of each possible permutation $\pi_i$ where $i = 1, \dots, D!$.
The probability distribution $P = \{p(\pi_1), \dots, p(\pi_{D!})\}$ is given by the relative frequencies:
\begin{equation}
    p(\pi_i) = \frac{\#\{s \mid s \leq N-(D-1)\tau; \, X_s^{(D, \tau)} \text{ has type } \pi_i\}}{N - (D-1)\tau}.
    \label{eq:supp_perm_prob}
\end{equation}
This distribution $P$ characterizes the ordinal properties of the time series.
For a purely random process (white noise), all $D!$ patterns are equally likely, resulting in a uniform distribution $P_e$.

\subsection{Permutation Entropy ($H$)}
The degree of disorder in the ordinal structure is quantified by the Shannon entropy of the distribution $P$.
To facilitate comparison across different embedding dimensions, we used the normalized permutation entropy $H[P]$:
\begin{equation}
    H[P] = \frac{S[P]}{S_{max}} = \frac{-\sum_{i=1}^{D!} p(\pi_i) \ln p(\pi_i)}{\ln(D!)},
\end{equation}
where $0 \leq H[P] \leq 1$.
\begin{itemize}
    \item $H \approx 1$ indicates maximum randomness (unpredictable ordinal sequences).
    \item $H \approx 0$ indicates a highly predictable, monotonic, or deterministic system.
\end{itemize}

\subsection{Statistical Complexity ($C$)}
Entropy alone cannot distinguish between systems with different levels of structural organization.
For instance, both a perfectly regular sequence and a perfectly random sequence lack non-trivial structure.
Statistical complexity $C[P]$~\cite{PhysRevLett.99.154102} measures the departure of the observed distribution $P$ from the uniform distribution $P_e$:
\begin{equation}
    C[P] = Q_J[P, P_e] \cdot H[P],
\end{equation}
where $Q_J[P, P_e]$ is the \textit{disequilibrium} measure~\cite{LAMBERTI2004119} based on the Jensen--Shannon divergence:
\begin{equation}
    Q_J[P, P_e] = Q_0 \left\{ S\left[ \frac{P + P_e}{2} \right] - \frac{1}{2}S[P] - \frac{1}{2}S[P_e] \right\}.
\end{equation}
The normalization constant $Q_0$ is the inverse of the maximum possible value of the Jensen--Shannon divergence, which occurs when $P$ is a delta distribution (all mass on one state):
\begin{equation}
    Q_0 = -2 \left\{ \frac{D!+1}{D!} \ln(D!+1) - 2 \ln(2D!) + \ln(D!) \right\}^{-1}.
\end{equation}
This ensures that $0 \leq C[P] \leq C_{max} < 1$.
In the CECP, $C[P]$ vanishes in the limits of perfect order ($H=0$) and white noise ($H=1$), thus identifying structured dynamics intermediate between these two states.

\section{Directed Horizontal Visibility Graph Analysis}
\label{sec:supp_hvg}

\subsection{Graph Construction and Degree Distributions}
For a time series $\{x_i\}_{i=1}^N$ of length $N$, horizontal visibility~\cite{Luque2009} between two observations $i$ and $j$ ($i<j$) is established when the following condition holds for every intermediate observation $n$ ($i<n<j$):
\begin{equation}
    x_i > x_n \quad \text{and} \quad x_j > x_n
\end{equation}
If this condition is satisfied, a directed edge is created from the earlier node to the later node.
Each node $i$ then has an in-degree $k_{\text{in}}$, associated with links arriving from the past, and an out-degree $k_{\text{out}}$, associated with links extending toward the future.
The time irreversibility of the series is quantified through the information-theoretic distance between the in-degree distribution $P(k_{\text{in}})$ and the out-degree distribution $P(k_{\text{out}})$ measured over the full series~\cite{Lacasa2012}.

\subsection{Kullback--Leibler and Jensen--Shannon Divergences for Irreversibility}
We use the Kullback--Leibler (KL) divergence to quantify asymmetry between the two distributions.
Writing $P_{\text{in}}(k)$ for $P(k_{\text{in}})$ and $P_{\text{out}}(k)$ for $P(k_{\text{out}})$ gives two directional KL divergences, where the sum over $k$ includes all degrees observed in the network:
\begin{equation}
    D_{\mathrm{KL}}(P_{\mathrm{in}} \Vert P_{\mathrm{out}}) = \sum_k P_{\mathrm{in}}(k) \ln \left( \frac{P_{\mathrm{in}}(k)}{P_{\mathrm{out}}(k)} \right)
\end{equation}
\begin{equation}
    D_{\mathrm{KL}}(P_{\mathrm{out}} \Vert P_{\mathrm{in}}) = \sum_k P_{\mathrm{out}}(k) \ln \left( \frac{P_{\mathrm{out}}(k)}{P_{\mathrm{in}}(k)} \right)
\end{equation}
In empirical time series, one distribution may assign zero probability to a degree $k$ for which the other distribution is nonzero, for example when $P_{\text{out}}(k)=0$ but $P_{\text{in}}(k)>0$.
This creates singularities in the denominator of the logarithmic term.
To reduce this instability and account for the asymmetry of the KL divergence, we used the symmetrized KL divergence ($D_{\mathrm{KL}}^{\mathrm{sym}}$) and, as the main distance measure, the more stable Jensen--Shannon divergence ($D_{\mathrm{JS}}$).
\begin{equation}
    D_{\mathrm{KL}}^{\mathrm{sym}} = D_{\mathrm{KL}}(P_{\mathrm{in}} \Vert P_{\mathrm{out}}) + D_{\mathrm{KL}}(P_{\mathrm{out}} \Vert P_{\mathrm{in}})
\end{equation}
\begin{equation}
    D_{\mathrm{JS}}(P_{\mathrm{in}} \Vert P_{\mathrm{out}}) = \frac{1}{2} D_{\mathrm{KL}}(P_{\mathrm{in}} \Vert M) + \frac{1}{2} D_{\mathrm{KL}}(P_{\mathrm{out}} \Vert M)
\end{equation}
Here, $M(k) = \frac{1}{2}[P_{\mathrm{in}}(k) + P_{\mathrm{out}}(k)]$ is the arithmetic mean of the two distributions.
Because the JS divergence uses this average distribution as the reference, it avoids zero-denominator singularities.
Values near zero indicate little mismatch in these degree marginals.
They do not establish reversibility of the full process.
Larger values describe stronger in/out-distribution mismatch, to be assessed against finite-sample surrogate references.

\subsection{Directional Asymmetry and High-Degree Restriction}
In addition to the magnitude measured by $D_{\mathrm{JS}}$, the orientation and degree localization of the in/out asymmetry are characterized by two complementary quantities.
The normalized signed KL imbalance,
\begin{equation}
    A_{\mathrm{KL}} = \frac{D_{\mathrm{KL}}(P_{\mathrm{in}}\Vert P_{\mathrm{out}}) - D_{\mathrm{KL}}(P_{\mathrm{out}}\Vert P_{\mathrm{in}})}{D_{\mathrm{KL}}(P_{\mathrm{in}}\Vert P_{\mathrm{out}}) + D_{\mathrm{KL}}(P_{\mathrm{out}}\Vert P_{\mathrm{in}})} \in [-1,1],
\end{equation}
isolates the direction of the mismatch from its magnitude.
The degree-resolved asymmetry,
\begin{equation}
    A(k) = \frac{P_{\mathrm{in}}(k) - P_{\mathrm{out}}(k)}{P_{\mathrm{in}}(k) + P_{\mathrm{out}}(k)},
\end{equation}
decomposes the signed mismatch across degrees, separating bulk (low-$k$) contributions from those at high-visibility (high-$k$) nodes.
Here $k$ is the number of visible past or future neighbors, not an elapsed-time horizon.
For example, $k=6$--10 does not denote a six- to ten-minute interval.

For the high-degree analysis, both $D_{\mathrm{JS}}$ and $A_{\mathrm{KL}}$ are recomputed after restricting the in- and out-degree histograms to $k \geq 10$.
The cutoff is an empirical operational boundary informed by the observed degree profiles, rather than a theoretically universal or prespecified threshold.
Degree measures visibility and need not uniquely identify an extreme-return event.
The restricted in- and out-degree histograms are separately normalized, so the resulting divergences compare conditional degree tails rather than an identical set of selected observations.
Consequently, the restricted $A_{\mathrm{KL}}$ is a nonlinear orientation summary of the entire renormalized tail and is not an arithmetic average of the displayed pointwise $A(k)$ values.
A positive $A(k)$ over part of the range can therefore coexist with a negative tail-restricted $A_{\mathrm{KL}}$ when the remaining tail bins have a different distributional orientation.
The same threshold is applied uniformly across all assets and surrogate ensembles to ensure a consistent comparison.

\section{Surrogate Data Generation}
\label{sec:supp_surrogates}
To quantitatively test nonlinear structure and correlations in the time series, we used Shuffle and Iterative Amplitude Adjusted Fourier Transform (IAAFT) surrogate data methods~\cite{Theiler1992, Schreiber1996}.
These methods serve as null models for identifying the origin of the observed dynamical properties by randomizing only selected statistical features of the original series.
For each asset, we generated 100 independent realizations per surrogate type, and reported the ensemble mean and standard deviation of each diagnostic across the realizations.

\subsection{Shuffle}
Shuffle surrogates are generated by randomly permuting the order of the original time series.
This operation removes all linear autocorrelation and nonlinear temporal dependence, while preserving the values of the individual data points.
The amplitude distribution and moments of the original series, including the mean, variance, skewness, and kurtosis, are therefore retained.
This null model retains the empirical distribution, including ties and finite-sample characteristics, after randomizing temporal order.
It provides a conditional reference for departures from an independently ordered sequence.

\subsection{Iterative Amplitude Adjusted Fourier Transform (IAAFT)}
IAAFT surrogates are used to randomize nonlinear temporal structure while preserving both the linear correlations and amplitude distribution of the original signal~\cite{Schreiber1996}.
Standard Fourier surrogates, which randomize only the phases, tend to produce Gaussian amplitude distributions after the inverse transform because of the central limit theorem.
For non-Gaussian data, this loss of higher-order moments can increase the risk of type-I errors.

The IAAFT algorithm alternates between frequency-space and time-space corrections until convergence.
Let the original time series of length $N$ be $\{x_n\}$, and let $S_x$ denote the sorted set of its values.
The initial sequence $\{x_n^{(0)}\}$ is generated by randomly shuffling the original series.

At the frequency-space correction step of iteration $i$, the Fourier phase $\phi_k^{(i-1)}$ of the current surrogate $\{x_n^{(i-1)}\}$ is retained, while the amplitude spectrum is replaced by the original spectrum $|X_k|$ to form new frequency coefficients:
\begin{equation}
    C_k^{(i)} = |X_k| e^{j \phi_k^{(i-1)}}.
\end{equation}
After inverse Fourier transformation, the temporary series $\{c_n^{(i)}\}$ is rank-ordered, and each element is replaced by the corresponding value from $S_x$ to obtain the new surrogate series $\{x_n^{(i)}\}$.

These two steps, spectral correction followed by amplitude replacement, are repeated until the relative change in power-spectrum error falls below the convergence tolerance ($10^{-6}$) or the maximum number of iterations (100) is reached.
The final surrogate reproduces the univariate amplitude distribution while approximately matching the power spectrum.
Reaching the iteration cap does not establish exact spectral convergence.
Shuffle and IAAFT comparisons identify departures from these constraints, but residuals can also reflect nonstationarity and do not uniquely identify a nonlinear generating mechanism.

\section{Post-2020 Cutoff Justification}
\label{sec:supp_scope}

\begin{figure}[pos=htbp]
    \centering
    \includegraphics[width=0.9\textwidth]{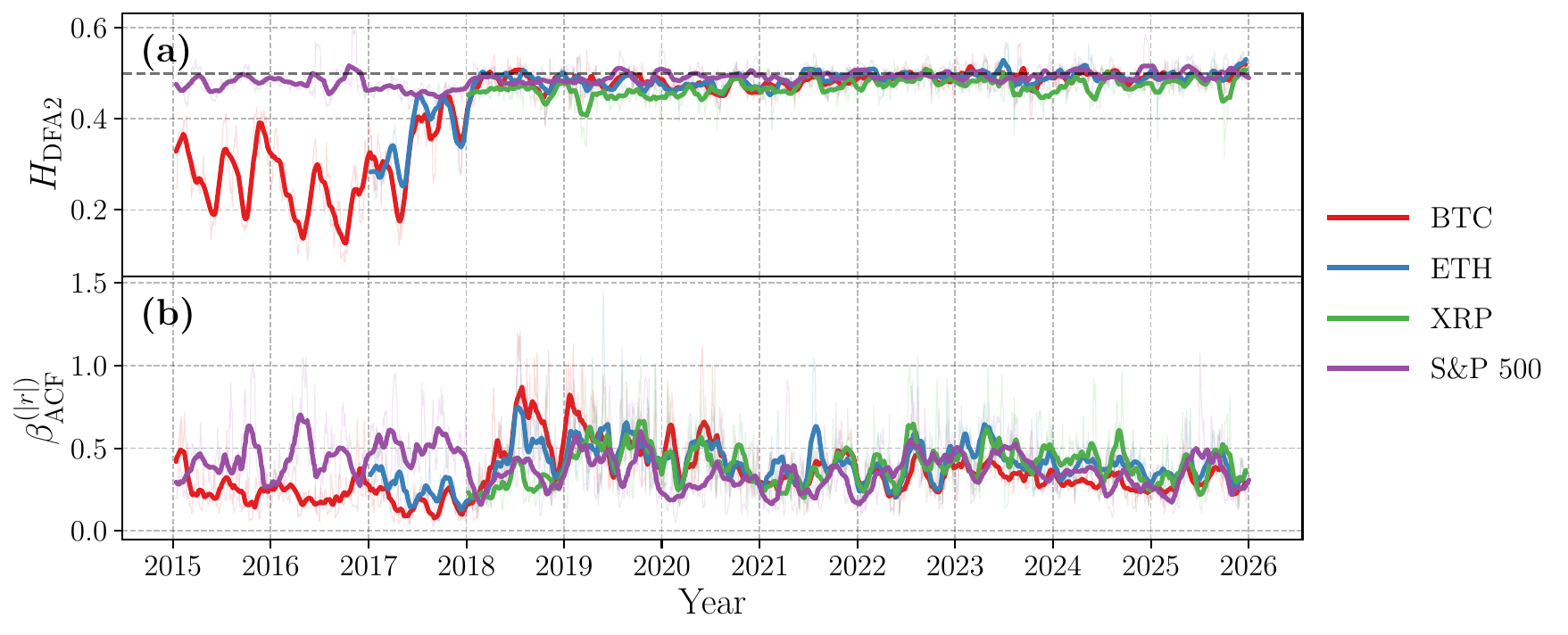}
    \caption{
        Rolling-window scaling diagnostics on deseasonalized minute returns, 2015--2025 ($W = 20{,}000$ minutes, centered moving average of length $30$).
        (a) Local DFA-2 Hurst exponent of returns (dashed line: random-walk benchmark $H = 0.5$).
        (b) Local volatility-ACF decay exponent.
        Faint and dark traces show unsmoothed estimates and centered moving averages, respectively.
        In both panels, cryptocurrency trajectories converge to the equity benchmark around 2019--2020, thereafter evolving within a common dispersion band.
    }
    \label{fig:supp_scope_rolling}
\end{figure}

\begin{figure}[pos=htbp]
    \centering
    \includegraphics[width=0.9\textwidth]{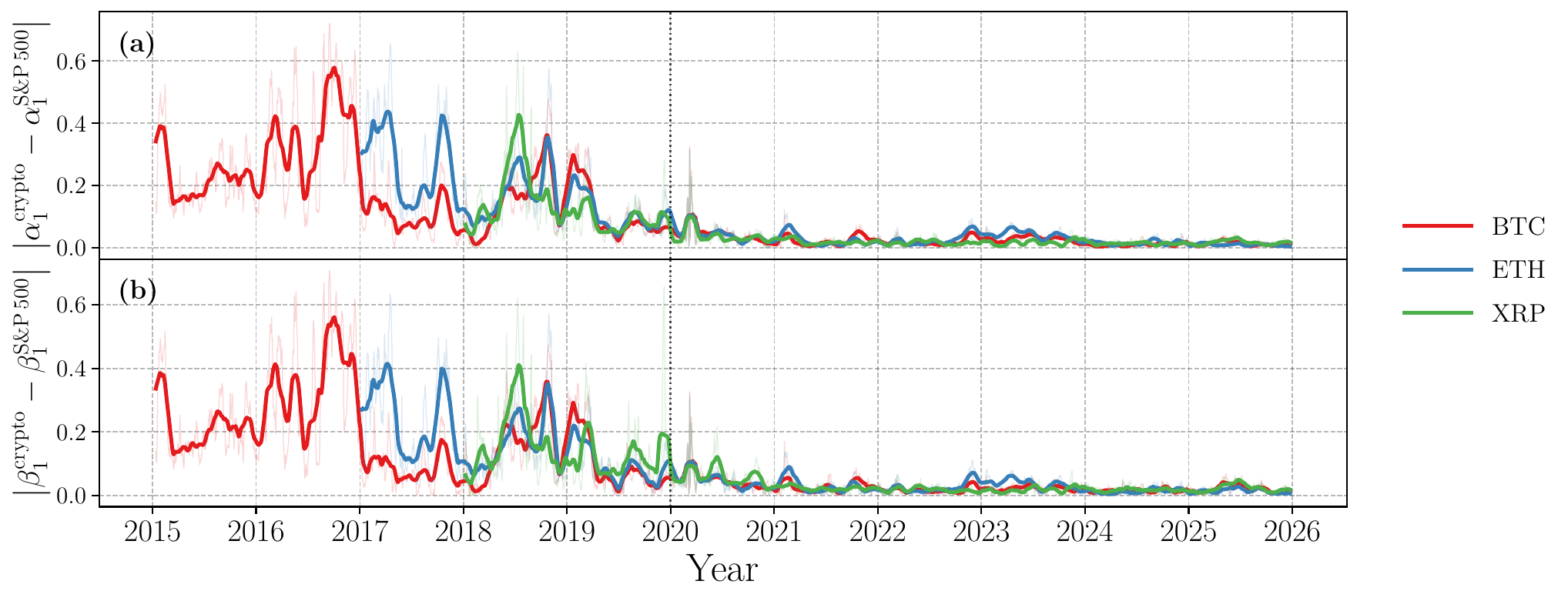}
    \caption{
        Rolling-window GARCH$(1,1)$ parameter distance from the S\&P~500 on deseasonalized minute returns, 2015--2025 ($W = 20{,}000$ minutes, moving average of length $30$).
        Faint traces show unsmoothed estimates.
        (a) Shock-sensitivity distance $|\alpha_1^{\text{crypto}} - \alpha_1^{\text{S\&P~500}}|$.
        (b) Persistence distance $|\beta_1^{\text{crypto}} - \beta_1^{\text{S\&P~500}}|$.
        Vertical dotted line marks January~1, 2020.
        Both distances drop to $\lesssim 0.05$ from 2020 onward, providing a parametric counterpart to the scaling diagnostics in Fig.~\ref{fig:supp_scope_rolling}.
    }
    \label{fig:supp_scope_garch}
\end{figure}

The post-2020 production window is chosen because the cryptocurrency series have reached a common mature-market regime under several independent rolling diagnostics by that date.
The exchange-level data coverage already reported in Table~\ref{tab:data_desc} (Sec.~\ref{sec:supp_data_construction}) shows that all three cryptocurrency assets are simultaneously covered across Bitstamp, Coinbase, and Kraken from March~1, 2019 onward (XRP's Coinbase listing, the latest of the nine exchange--asset onboarding dates).
Figs.~\ref{fig:supp_scope_rolling}--\ref{fig:supp_scope_garch} then show the empirical convergence that motivates excluding the earlier immature-market period from the quantitative comparisons in the main text.

The main text restricts the comparative analysis to the period from January~1, 2020 to December~31, 2025.
This cutoff combines an institutional rationale (post-2020 covers the period in which BTC, ETH, and XRP are simultaneously available on all three cryptocurrency exchanges in our universe and includes the subsequent spot Bitcoin ETF approval and large-cap institutional participation) with the empirical convergence diagnostics described in this section.

We compute three rolling diagnostics on deseasonalized minute returns over the 2015--2025 sample.
Each window spans $W = 20{,}000$ consecutive minutes (approximately fourteen trading days).
The per-window series are smoothed by a centered moving average of length $30$ to suppress sub-window fluctuations.
The diagnostics are the local DFA-2 Hurst exponent of returns~\cite{PhysRevE.49.1685} and the local volatility-ACF decay exponent, shown in Figure~\ref{fig:supp_scope_rolling}, panels~(a) and~(b) respectively.
The third is the rolling GARCH$(1,1)$ parameter distance~\cite{Bollerslev1986} from the S\&P~500, shown in Figure~\ref{fig:supp_scope_garch}.
The first two are nonparametric scaling diagnostics.
The third places each cryptocurrency in the GARCH parameter space spanned by $(\alpha_1, \beta_1)$ relative to the equity benchmark fitted on the same rolling window.

The local DFA-2 Hurst exponent of returns shows that BTC, ETH, and XRP exhibit a markedly anti-persistent regime ($H \lesssim 0.4$) with strong inter-asset dispersion through 2017--2018, which then drifts upward and stabilizes near the random-walk benchmark ($H \approx 0.5$) by 2019--2020.
Within this transition, XRP is the lagging asset: BTC and ETH reach the $H \approx 0.5$ band by mid-2018, but XRP remains visibly below ($H \approx 0.42\text{--}0.47$) through 2018--2019 and only joins the common band in late 2019 to early 2020.
After roughly 2020 the four DFA Hurst trajectories---including the equity benchmark, which sits near $H \approx 0.5$ throughout---evolve in a common narrow band, with no systematic asset-class separation.
The volatility-ACF decay exponent shows the same qualitative transition: the early-period spread between cryptocurrencies and the equity benchmark collapses around 2019--2020, after which all four trajectories share a comparable mean level and dispersion envelope.

Figure~\ref{fig:supp_scope_garch} shows the rolling GARCH parameter distance.
For each cryptocurrency $C \in \{\text{BTC}, \text{ETH}, \text{XRP}\}$, we fit a GARCH$(1,1)$ model~\cite{Bollerslev1986} with Student-$t$ errors on every rolling window and record $|\alpha_1^{C} - \alpha_1^{\text{S\&P~500}}|$ and $|\beta_1^{C} - \beta_1^{\text{S\&P~500}}|$, where $\alpha_1^{\text{S\&P~500}}, \beta_1^{\text{S\&P~500}}$ are obtained from the S\&P~500 fit on the same window.
Both distances are large and noisy through 2015--2018: $|\Delta \alpha_1|$ reaches $0.3\text{--}0.6$ for BTC over 2016--2017 and for ETH over 2017--2018, and $|\Delta \beta_1|$ shows comparable magnitudes.
A sharp contraction occurs over 2018--2019 across both panels, and from 2020 onward all three distances drop into a narrow band $\lesssim 0.05$ for the remainder of the sample, with only sporadic small excursions.
The convergence is thus reflected in both the nonparametric scaling diagnostics and the parameter space of the GARCH model.
We emphasise that the distances reported here are between observed GARCH parameters and are used only to justify the post-2020 scope.
We do not use GARCH surrogate decompositions as primary structural evidence, because GARCH coefficients explicitly parameterize temporal volatility dependence and destructive temporal nulls are not directly comparable to the ordinal and topological surrogate diagnostics in the main text.

The three rolling diagnostics consistently show that cryptocurrency trajectories converge to the equity benchmark's dispersion band around 2019--2020, with XRP as the final asset to transition.
The cutoff at January~1, 2020 therefore corresponds to the date by which the slowest-converging cryptocurrency in our basket has reached this common regime under all three diagnostics, rather than to a round-number choice.
Earlier years (2015--2019) instead reflect a cryptocurrency-side immature-market regime that dominates several stylized-fact comparisons and is not the dynamical regime of present economic interest.
The full sample is therefore used only as motivation for the cutoff and not as the basis of any quantitative claim in the main text.

\section{Post-2022 Re-analysis}
\label{sec:supp_post2022}

The main text restricts the sample to the post-2020 period on the basis of the rolling-window diagnostics presented in Sec.~\ref{sec:supp_scope}.
A natural concern is that the central conclusions---macroscopic similarity with residual CECP/HVG class structure---might be sensitive to the exact placement of this cutoff.
We therefore repeated the main deseasonalized diagnostics on a strictly later window starting January~1, 2022, retaining all other configuration choices.
The post-2022 re-analysis covers a $\sim 4$-year window, so a moderate increase in surrogate-ensemble sampling noise is expected.
MF-DFA~\cite{Kantelhardt_2002} is reported here as a sensitivity check on multiscale overlap, whereas CECP and HVG carry the residual class-structure claim.

\subsection{Upper-tail power-law slope}
\label{subsec:supp_tail}

Table~\ref{tab:supp_tail_post2022} reports the same deseasonalized MLE-based power-law slope diagnostic~\cite{Clauset_2009} used in the main text, but with the sample restricted to observations after January~1, 2022.
The post-2022 estimates remain in the heavy-tailed regime for all four assets ($\alpha < 4$).
The S\&P~500 tail slope increases relative to the post-2020 main-text estimate and moves closer to the inverse-cubic benchmark~\cite{PhysRevE.60.5305}, illustrating that the estimated power-law exponent is sensitive to the selected sample window and tail curvature.
For this reason, both the main text and this robustness check use $\alpha$ only as a macroscopic distributional diagnostic, not as evidence for asset ranking, temporal convergence, or crypto--equity class separation.

\begin{table}[pos=htbp]
    \centering
    \setlength{\tabcolsep}{12pt}
    \renewcommand{\arraystretch}{0.85}
    \caption{
        Post-2022 deseasonalized upper-tail power-law slope estimates.
        The estimates use the same KS-selected $x_{\min}$ maximum-likelihood procedure as the main-text tail analysis.
    }
    \label{tab:supp_tail_post2022}
    \begin{tabular}{lcccc}
        \toprule
        Asset    & $\alpha$ & SE      & $k$        & $x_{\min}$ \\
        \midrule
        BTC      & $3.053$  & $0.033$ & $8{,}365$  & $4.9163$   \\
        ETH      & $2.976$  & $0.031$ & $9{,}523$  & $4.6768$   \\
        XRP      & $2.820$  & $0.021$ & $17{,}505$ & $3.6849$   \\
        S\&P~500 & $3.106$  & $0.031$ & $9{,}830$  & $4.0697$   \\
        \bottomrule
    \end{tabular}
\end{table}

\subsection{MF-DFA multifractal spectrum width}
\label{subsec:supp_mfdfa}

\begin{figure}[pos=htbp]
    \centering
    \includegraphics[width=0.8\textwidth]{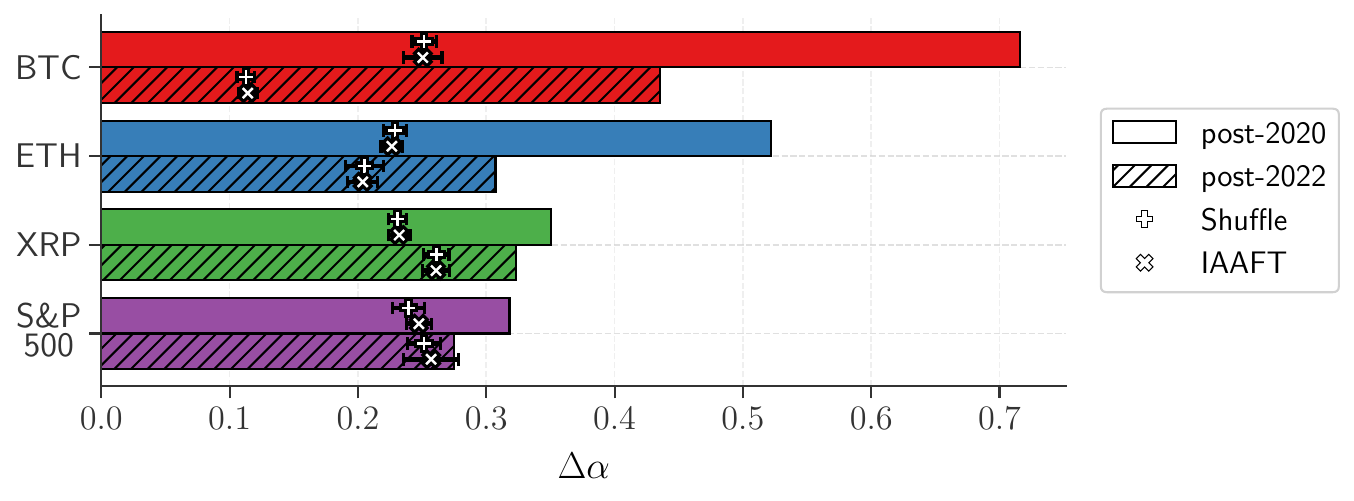}
    \caption{
        Robustness of the MF-DFA multifractal spectrum width $\Delta\alpha$ to the start-date cutoff.
        Solid bars show the post-2020 window and hatched bars show the post-2022 window.
        Surrogate ensemble results are overlaid as markers with error bars (plus markers for Shuffle and cross markers for IAAFT, showing mean $\pm$ s.d.\ over 100 realizations).
        Bitcoin retains the broadest spectrum in both windows, but the ordering among ETH, XRP, and S\&P~500 is not preserved: Ethereum's spectrum narrows the most, so XRP becomes the second-widest in the post-2022 window.
        The shuffle/IAAFT equivalence is preserved across both windows.
    }
    \label{fig:supp_mfdfa}
\end{figure}

Figure~\ref{fig:supp_mfdfa} compares the multifractal spectrum width $\Delta\alpha$ of the original returns and of the shuffle and IAAFT surrogate ensembles between the post-2020 and post-2022 windows.
We use this comparison as a sensitivity check rather than as a class-discrimination result.

First, the observed $\Delta\alpha$ values remain broadly comparable across cutoffs, with BTC retaining the broadest spectrum in both windows.
The absolute magnitudes shrink in the post-2022 sample (BTC $0.72 \to 0.44$, ETH $0.52 \to 0.31$, XRP $0.35 \to 0.32$, S\&P~500 $0.32 \to 0.27$), consistent with the narrower observation window providing fewer extreme cascades.
Because this ordering is not aligned with a clean crypto--equity split, we treat it as a sensitivity pattern rather than class evidence.

Second, the shuffle and IAAFT ensembles agree to within their per-realization standard deviations in both windows ($\Delta\alpha_{\mathrm{shuf}} \approx \Delta\alpha_{\mathrm{IAAFT}}$ for every ticker).
This supports the conservative main-text use of MF-DFA: the multiscale profiles and surrogate controls are useful macroscopic sensitivity diagnostics, but $\Delta\alpha_{\rm surr}$ is not used as a crypto--equity class discriminator.

\subsection{Complexity--entropy causality plane}
\label{subsec:supp_cecp}

\begin{figure}[pos=htbp]
    \centering
    \includegraphics[width=0.85\textwidth]{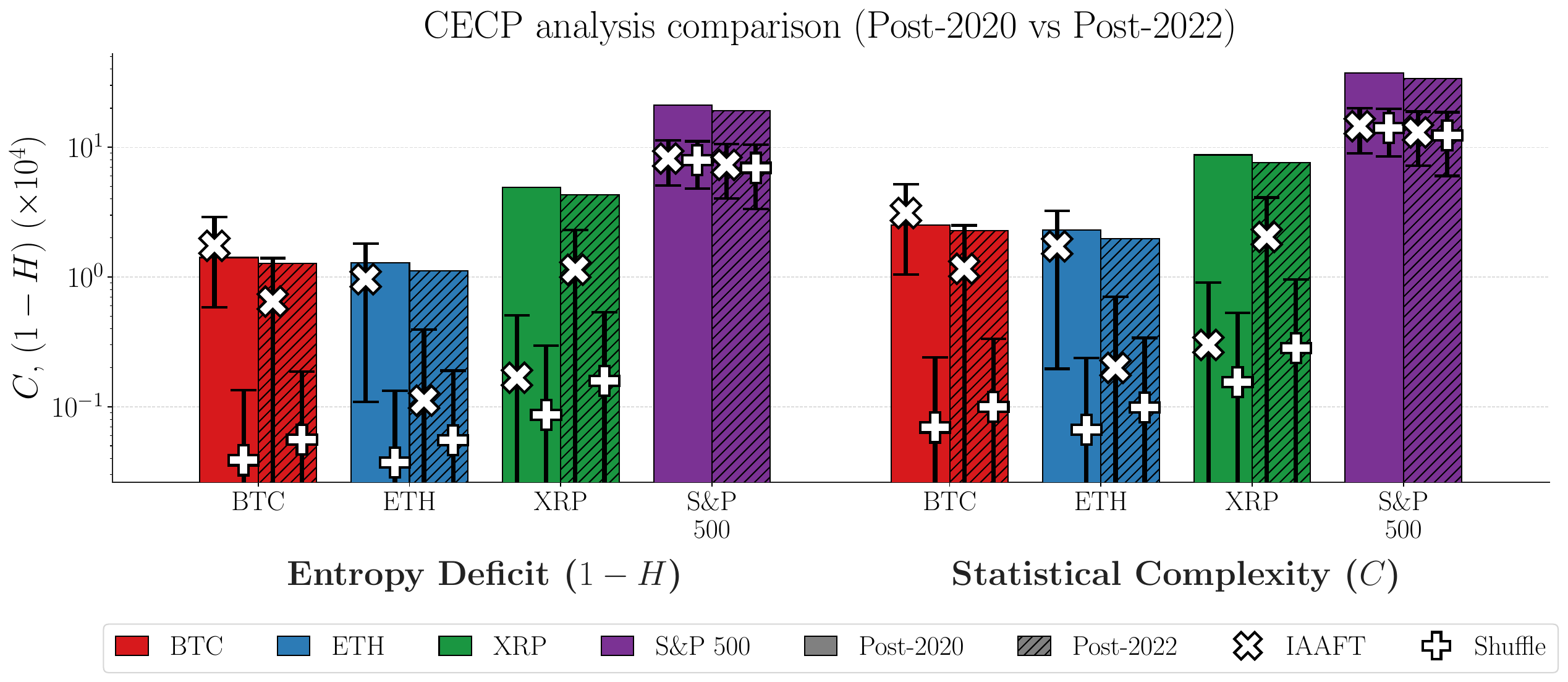}
    \caption{
        Robustness of the complexity--entropy causality plane (CECP) coordinates to the start-date cutoff.
        The figure compares the original and surrogate coordinates ($1-H$ in the left panel, $C$ in the right panel) between the post-2020 (solid bars) and post-2022 (hatched bars) windows.
        Original returns are plotted as bars, while surrogate ensemble results are overlaid as markers with error bars (cross markers for IAAFT and plus markers for Shuffle, showing mean $\pm$ s.d.\ over 100 realizations).
        All values are scaled by $10^4$.
        The cross-asset hierarchy and the relationships between the empirical data and the surrogates are consistent across both analysis windows.
    }
    \label{fig:supp_cecp}
\end{figure}

Figure~\ref{fig:supp_cecp} compares the CECP coordinates, namely the entropy deficit $1-H$ and the statistical complexity $C$, between the post-2020 and post-2022 windows.
These coordinates, defined mathematically in Sec.~\ref{sec:supp_cecp}, are plotted for both the original return series and their surrogate ensembles.
In both windows, the normalized permutation entropy $H$ is extremely close to unity for all three cryptocurrencies, keeping the entropy deficit $1-H$ below $10 \times 10^{-4}$.
In comparison, the S\&P~500 futures exhibit a larger entropy deficit of approximately $20 \times 10^{-4}$.
The statistical complexity $C$ shows a similar level of stability across both sample periods, ensuring that the qualitative hierarchy
\begin{equation}
    C_{\text{S\&P}} \;\gg\; C_{\text{XRP}} \;>\; C_{\text{BTC}} \;\approx\; C_{\text{ETH}}
    \label{eq:supp_cecp_order}
\end{equation}
is preserved.
The S\&P~500 sits at lower $H$ and higher $C$ than any cryptocurrency in both windows, reproducing the main-text claim that the equity benchmark retains additional ordinal structure.

Shuffling the return series to destroy all temporal dependencies causes the cryptocurrency coordinates to collapse close to the random limit of zero complexity and zero entropy deficit.
This confirms that their original spatial displacement is driven by temporal correlations rather than marginal distribution effects.
The relative behavior under the IAAFT surrogate control is similarly preserved.
For BTC and ETH, the linear correlations preserved by the IAAFT surrogates explain most of their original displacement.
In contrast, XRP's coordinates collapse close to the shuffled baseline, which highlights a dominantly nonlinear temporal structure.
Finally, the S\&P~500 retains a large original--surrogate gap in both periods, showing that its structured ordinal organization remains robust to the start-date cutoff.
At the same time, the S\&P~500's high surrogate complexity floor—driven by the concentration of zero returns—is similarly preserved across both windows.

\subsection{HVG directional asymmetry}
\label{subsec:supp_hvg}

\begin{figure}[pos=htbp]
    \centering
    \includegraphics[width=\textwidth]{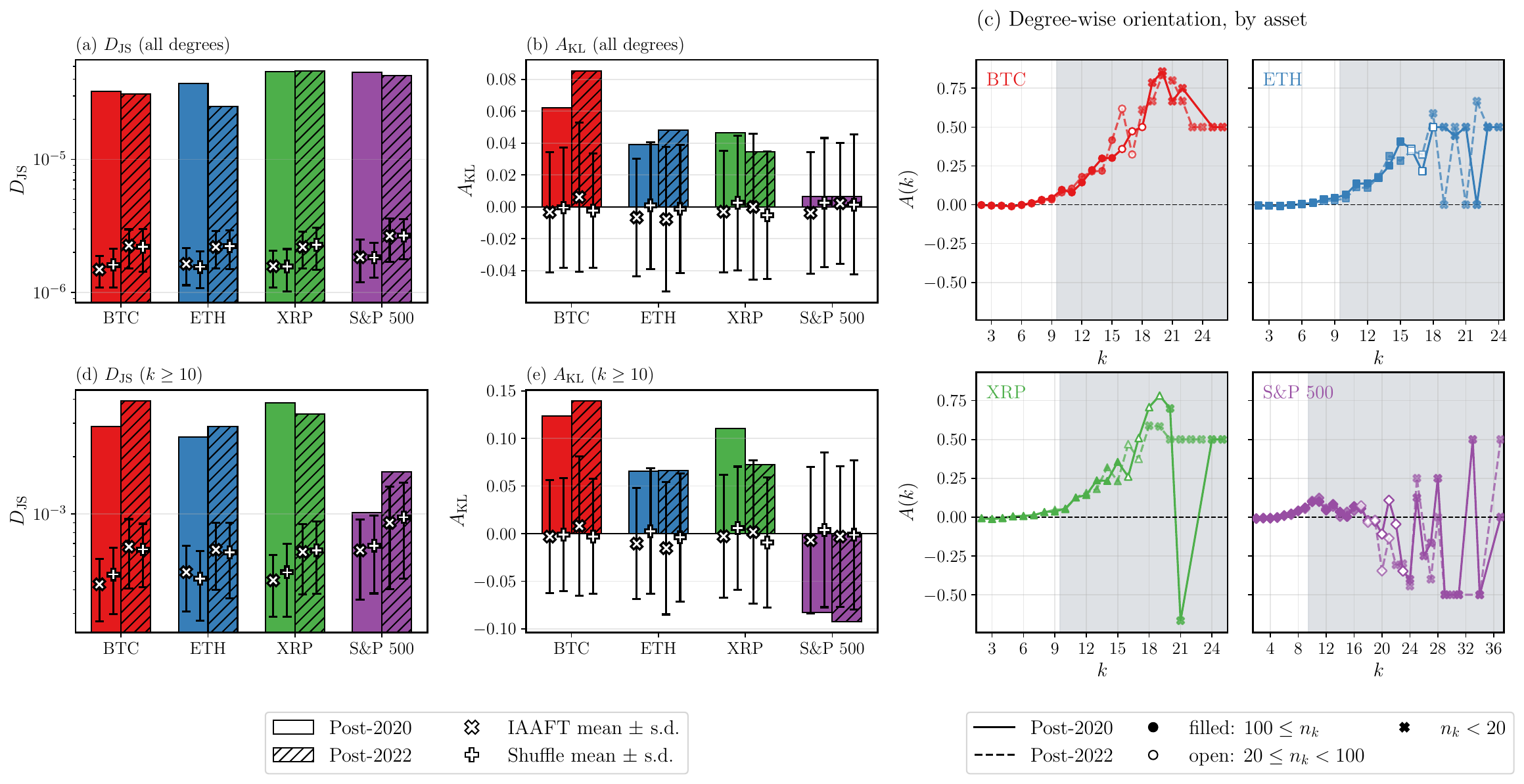}
    \caption{
        Comparison of signed-return HVG directional-asymmetry diagnostics between the post-2020 and post-2022 windows.
        Diagnostics are computed over all degrees for (a) $D_{\mathrm{JS}}$ (log scale) and (b) $A_{\mathrm{KL}}$ (linear scale), and restricted to the high-degree sector $k\geq10$ in (d, e).
        Panel (c) shows the degree-wise asymmetry profile $A(k)$ for each asset separately, with the post-2020 window as a solid line and the post-2022 window as a dashed line; the gray band marks the $k\geq10$ sector used in (d, e), and all four subpanels share a common $A(k)$ axis.
        For panels (a, b, d, e), the post-2020 window is shown as solid (plain) bars and the post-2022 window as hatched bars.
        Marker style encodes the per-degree support $n_k=c_\mathrm{in}(k)+c_\mathrm{out}(k)$, independent of window: filled markers denote $n_k\geq100$, open markers $20\leq n_k<100$, and $\times$ markers $n_k<20$.
        In (a, b, d, e), open crosses (IAAFT) and plus markers (Shuffle) indicate surrogate ensemble means $\pm$ one standard deviation.
    }
    \label{fig:supp_hvg}
\end{figure}

Figure~\ref{fig:supp_hvg} compares the Jensen--Shannon divergence $D_{\mathrm{JS}}$ and the directional asymmetry $A_{\mathrm{KL}}$ between the in- and out-degree distributions of the directed HVG.
These directional network metrics, defined mathematically in Sec.~\ref{sec:supp_hvg}, are plotted alongside the surrogate ensembles detailed in Sec.~\ref{sec:supp_surrogates} for both the post-2020 and post-2022 windows.
The all-degree divergence and the high-degree crypto--equity contrast are qualitatively unchanged under the later cutoff.
The post-2022 surrogate ensembles have wider dispersion in this comparison.
The change in sample length occurs together with a change in the market periods represented.
The high-degree $A_\mathrm{KL}$ sign pattern is preserved---positive for the three cryptocurrencies and negative for the S\&P~500.

\section{CECP Parameter Sensitivity}
\label{sec:supp_cecp_grid}

The CECP separation reported in the main text is stable across a broad short-memory embedding grid.
The main CECP analysis uses embedding dimension $m=5$ and delay $\tau=1$.
To verify that the crypto--equity separation is not an artifact of this production choice, we recomputed the CECP coordinates over an embedding grid
\[
    m \in \{3,4,5,6,7\}, \qquad \tau \in \{1,2,3,5,10\}.
\]
The grid search was performed for both the post-2020 production window and the post-2022 robustness window, using the same deseasonalized returns and surrogate settings as in the main CECP analysis.

\begin{figure}[pos=htbp]
    \centering
    \includegraphics[width=\textwidth]{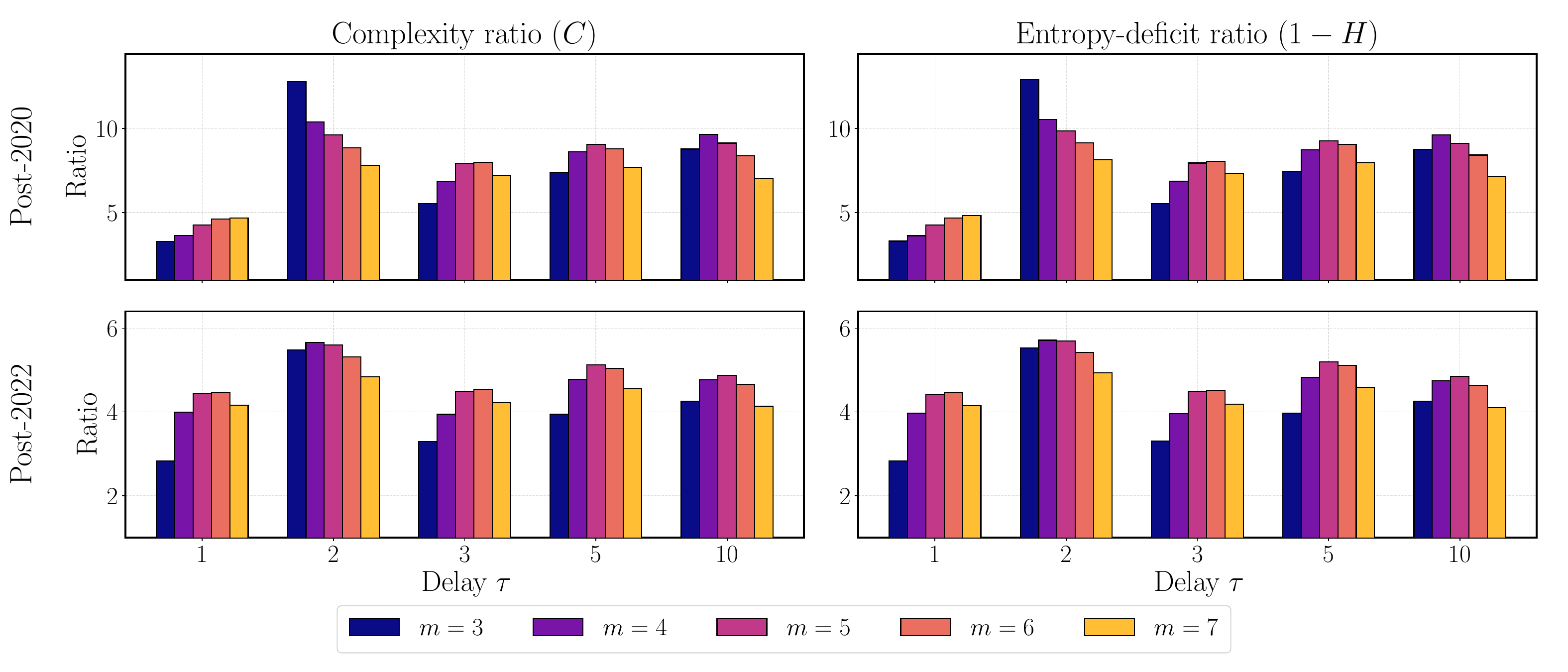}
    \caption{
    Robustness of the CECP crypto--equity separation to embedding dimension $m$ and delay $\tau$, shown as grouped bars for each $m$ as a function of delay $\tau$.
    Left column: statistical-complexity ratio $C_{\mathrm{S\&P}}/\max(C_{\mathrm{crypto}})$.
    Right column: entropy-deficit ratio $(1-H)_{\mathrm{S\&P}}/\max(1-H)_{\mathrm{crypto}}$.
    Top row: post-2020 production window.
    Bottom row: post-2022 robustness window.
    All bars exceed one, indicating that the S\&P~500 remains lower-entropy and higher-complexity than all three cryptocurrencies throughout the tested embedding grid.
    }
    \label{fig:supp_cecp_grid}
\end{figure}

Figure~\ref{fig:supp_cecp_grid} summarizes the separation using two ratios: the statistical-complexity ratio $C_{\mathrm{S\&P}}/\max(C_{\mathrm{crypto}})$ and the entropy-deficit ratio $(1-H)_{\mathrm{S\&P}}/\max(1-H)_{\mathrm{crypto}}$.
Values above one mean that the S\&P~500 has higher complexity, or a larger departure from maximum permutation entropy, than every cryptocurrency at the same $(m,\tau)$.
In the post-2020 window, the S\&P~500 has the highest $C$ and the largest entropy deficit across all 25 tested parameter combinations, and the same holds in the post-2022 window.
The production setting $(m,\tau)=(5,1)$ sits near the grid minimum in the post-2020 window; in the post-2022 window it lies below the grid median without approaching either extreme (Fig.~\ref{fig:supp_cecp_grid}).
The ratio rises at longer delays for most embedding dimensions, reaching its grid maximum at $\tau=2$ (at the lowest tested $m$ for post-2020, and one step higher in $m$ for post-2022), while the dependence on $m$ alone at fixed $\tau$ is comparatively mild.
Since $(m,\tau)=(5,1)$ never approaches the grid maximum, the main-text separation is, if anything, a conservative estimate.

\section{Degree-Resolved HVG Irreversibility}
\label{sec:supp_hvg_degree}

\begin{figure}[pos=htb]
    \centering
    \includegraphics[width=\textwidth]{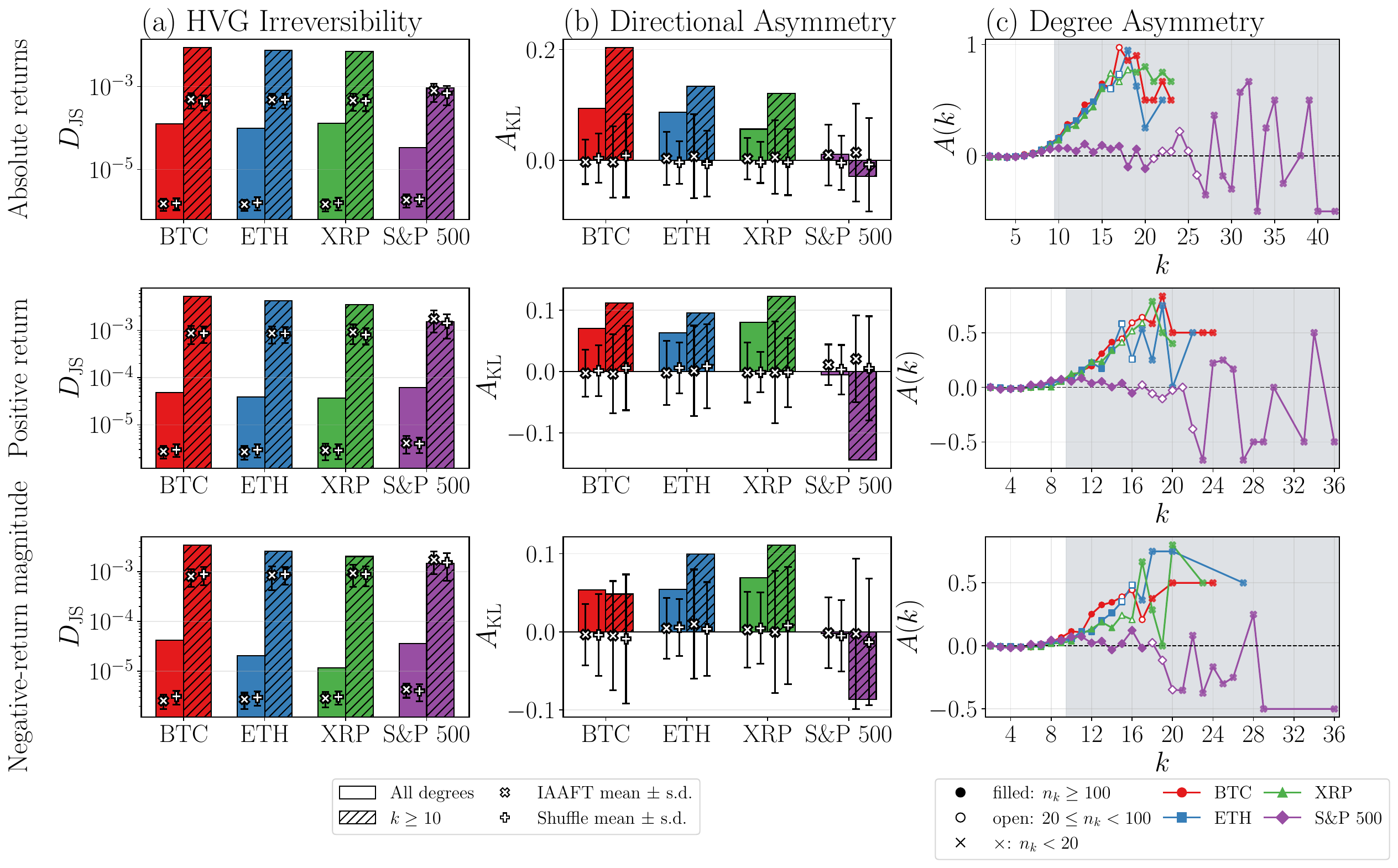}
    \caption{
        HVG irreversibility across return transformations in the post-2020 window.
        Rows correspond to the absolute-return series, positive-return subsequence, and negative-return-magnitude subsequence.
        The first two columns report the Jensen--Shannon divergence $D_\mathrm{JS}$ and signed directional asymmetry $A_\mathrm{KL}$ for all degrees and for the high-degree sector $k\geq10$.
        Surrogate-ensemble means are overlaid with $\pm$ s.d.\ error bars.
        The right column shows the local degree profile $A(k)$.
        Filled markers denote $n_k\geq100$, open markers denote $20\leq n_k<100$, and crosses connected by dashed lines denote $n_k<20$.
    }
    \label{fig:supp_hvg_transform_grid}
\end{figure}

The final robustness check asks which parts of the HVG degree distribution carry the crypto--equity separation.
Figure~\ref{fig:supp_hvg_transform_grid} extends the main-text HVG visualization from the raw signed-return series to the absolute-return series, the positive-return subsequence, and the negative-return-magnitude subsequence.
All three are evaluated over the primary post-2020 production window (January~1, 2020 to December~31, 2025) using the directed HVG structure defined in Sec.~\ref{sec:supp_hvg}.
The threshold $k=10$ is used as the same operational high-visibility cutoff as in the main text: it removes the bulk of the degree distribution while retaining enough observations for stable truncated-distribution estimates across all assets.
For each cryptocurrency and each return transform, we tested whether the crypto in-degree share $c_\mathrm{in}(k)/(c_\mathrm{in}(k)+c_\mathrm{out}(k))$ exceeds the corresponding S\&P~500 value using a one-sided Fisher exact test on the $2\times2$ table of in/out counts.
Only degrees for which both assets satisfy $c_\mathrm{in}(k)+c_\mathrm{out}(k)\geq100$ were tested, and $p$-values were corrected by Benjamini--Hochberg FDR~\cite{Benjamini1995} within each crypto--S\&P--500 pair.
This correction is local to the degree-wise pair comparisons and does not cover all transformations or diagnostics.
Because in/out counts come from the same temporally dependent graph, these tests are treated as exploratory localization aids rather than independent confirmatory evidence.
The first isolated significant degree can occur at small $k$, but the first consecutive significant run is more informative for identifying a stable separation region.
For the final price-derived, source-gap-pooled signed returns, the first two-degree significant run begins at $k=12$ for Bitcoin and XRP and at $k=14$ for Ethereum.
Thus $k\geq10$ should be read as an operational cutoff that includes the onset of the stable crypto--equity separation, not as a theoretically universal threshold.

Restricting the distributions to $k\geq10$ raises both the observed divergence and the surrogate values.
This increase in both observed and surrogate values means that high-degree irreversibility is not attributable to nonlinear temporal structure alone; marginal-distribution effects, finite sampling in the degree tail, and the linear spectral structure retained by IAAFT can also contribute.
Across all three return transformations, the high-degree $D_\mathrm{JS}$ of the S\&P~500 is largely accounted for by both the Shuffle and IAAFT surrogate ensembles, whereas the corresponding cryptocurrency values remain well above their surrogate ranges and are only weakly accounted for by either null model.
For the IAAFT comparison, the unadjusted $p_\mathrm{MC}$ remains at $1/101$ for BTC and ETH under all three transforms.
XRP gives $1/101$ for absolute returns and for the positive-return subsequence, and $3/101$ for negative-return magnitudes, while the S\&P~500 is not below 0.05 for any transform.
These results reinforce $D_\mathrm{JS}$ as the primary diagnostic: the surrogate-explainable S\&P~500 contrast is consistently different from the unexplained cryptocurrency contrast across transformations.
The full post-2020 transformation-family audit applies Holm adjustment across 128 combinations of asset, return transformation, degree sector, diagnostic, and null model; no comparison survives that correction.
The displayed $p_\mathrm{MC}$ values remain unadjusted within the corresponding tables, so the transformation result is interpreted as a specification-dependent pattern rather than a universal claim.

The high-degree $A_\mathrm{KL}$ values show a qualitatively consistent directional trend across the three transformations: they are positive for the cryptocurrencies and negative for the S\&P~500.
Because the surrogate comparisons for $A_\mathrm{KL}$ are not significant, this sign and magnitude pattern is treated only as qualitative directional evidence, not as a statistically established separation or a mechanistic claim.

The degree-wise Fisher tests provide pairwise comparisons between each cryptocurrency and the S\&P~500.
In the absolute-return series, the first significant run of two consecutive degrees begins at $k=8$ for Bitcoin and XRP, and at $k=9$ for Ethereum.
Within the positive-return subsequences, a consecutive high-degree run begins at $k=12$ for Bitcoin.
Ethereum has no consecutive run, while XRP has a low-degree run at $k=3$--4 and a separate high-degree run at $k=12$--13.
For negative-return magnitudes, a consecutive run begins at $k=12$ for Bitcoin and at $k=3$ for XRP.
Ethereum has only an isolated significant degree, at $k=14$.
The localization therefore varies materially with the return transform and should not be read as a universal degree threshold.
These degree-wise cross-asset tests are distinct from the surrogate comparisons in the first two columns, which evaluate whether the temporal structure within individual assets is replicable by their respective null models.

\subsection{Threshold and Pseudocount Sensitivity}

The threshold and pseudocount sensitivity check evaluates the primary price-derived signed-return diagnostics over the full grid of high-degree thresholds $K=8$--$12$ and pseudocounts $0.1$ and $0.5$.
For each asset and null model, Table~\ref{tab:price_contiguous_k_grid} reports the minimum and maximum unadjusted inclusive Monte Carlo $p$-values across the 10 threshold--pseudocount combinations.
This aggregation tests whether the qualitative surrogate-relative conclusions depend on either specification choice.

\begin{table}[htbp]
    \centering
    \caption{
        Threshold and pseudocount sensitivity of the primary price-derived signed-return diagnostics over all combinations of thresholds $K=8,\ldots,12$ and pseudocounts $0.1$ and $0.5$.
        The reported ranges represent the minimum and maximum unadjusted inclusive Monte Carlo $p$-values obtained across these 10 parameter settings.
    }
    \label{tab:price_contiguous_k_grid}
    \begin{tabular}{llrr}
        \toprule
        Asset & Null & $D_{\rm JS}$ $p$ range & $A_{\rm KL}$ $p$ range \\
        \midrule
           BTC   & SHUFFLE & 0.010--0.010 & 0.040--0.178 \\
           BTC   & IAAFT   & 0.010--0.010 & 0.079--0.158 \\
           ETH   & SHUFFLE & 0.010--0.010 & 0.198--0.535 \\
           ETH   & IAAFT   & 0.010--0.010 & 0.139--0.317 \\
           XRP   & SHUFFLE & 0.010--0.010 & 0.079--0.436 \\
           XRP   & IAAFT   & 0.010--0.010 & 0.040--0.416 \\
        S\&P~500 & SHUFFLE & 0.010--0.822 & 0.257--0.455 \\
        S\&P~500 & IAAFT   & 0.020--0.743 & 0.297--0.653 \\
        \bottomrule
    \end{tabular}
\end{table}

Across this grid, the cryptocurrency $D_\mathrm{JS}$ comparisons remain below 0.05 for both null models, while the S\&P~500 ranges are broader and include non-significant values.
The $A_\mathrm{KL}$ comparisons are not uniformly significant across the grid, so we retain $D_\mathrm{JS}$ as the primary high-degree diagnostic and interpret $A_\mathrm{KL}$ only as directional evidence.

\section{Temporal Aggregation Sensitivity}
\label{sec:revision_sampling}

Five- and 10-minute series retain only complete nonoverlapping buckets within the same timestamp segments.
Matched 100-Shuffle and 100-IAAFT ensembles use the same construction.
We restrict the reported aggregation range to intervals for which every asset retains at least one high-degree bin with $n_k\geq100$.

\subsection{Directed Horizontal Visibility Graph}
\label{subsec:supp_hvg_sampling}

\begin{figure}[pos=htbp]
    \centering
    \includegraphics[width=\textwidth]{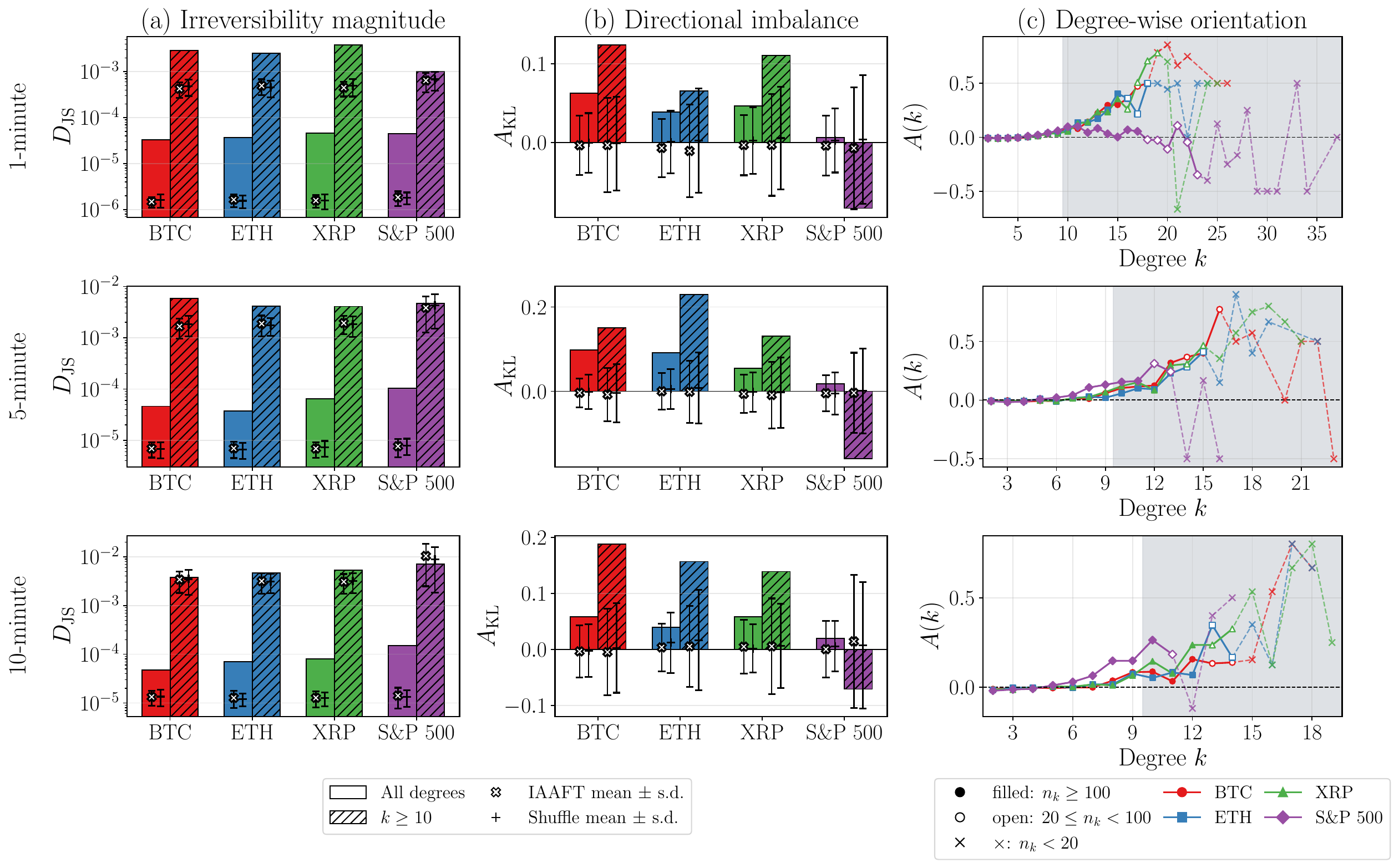}
    \caption{
        Sampling-interval sensitivity of the signed-return DHVG diagnostics.
        DHVGs are constructed separately within contiguous source-timestamp segments and only their degree counts are pooled, so no edge crosses a source gap.
        Rows show 1-, 5-, and 10-minute returns.
        Columns show $D_\mathrm{JS}$, $A_\mathrm{KL}$, and $A(k)$.
        Solid and hatched bars denote observed all-degree and $k\geq10$ values, respectively.
        Open crosses and plus markers show IAAFT and Shuffle means with one-standard-deviation error bars.
        The right column displays every degree with nonzero empirical support.
        Filled markers and solid lines denote $n_k=c_\mathrm{in}(k)+c_\mathrm{out}(k)\geq100$, open markers remain connected by solid lines for $20\leq n_k<100$, and crosses connected by dashed lines denote $n_k<20$.
        The last category is included only to reveal the sparsely sampled tail and is not interpreted as robust degree-wise structure.
    }
    \label{fig:supp_hvg_sampling}
\end{figure}

At one minute, the IAAFT high-degree $D_\mathrm{JS}$ comparison gives $p_\mathrm{MC}=0.010$ for each cryptocurrency and $0.109$ for the S\&P~500.
At five minutes, all three cryptocurrencies remain below 0.05 (BTC: 0.010, ETH: 0.030, XRP: 0.010), while the S\&P~500 does not (0.327).
At 10 minutes, none of the four comparisons is below 0.05.
Thus the surrogate-relative crypto separation weakens over the reported aggregation range, even though the observed metric magnitudes need not decrease monotonically.
Because coarser aggregation combines several short-horizon fluctuations into longer return increments, the interval comparison mixes the loss of high-frequency temporal detail with changes in the physical horizon and effective degree-tail support.
We therefore interpret the result as a high-frequency qualification rather than attributing it to a single aggregation mechanism.
The degree-resolved profiles further qualify this interpretation.
At one minute, $A(k)$ begins to rise from near zero around $k=6$--10 for all four assets, including the S\&P~500, so the onset of the moderate-degree imbalance is not a crypto-specific signature.
The cryptocurrency profiles then continue to increase into higher, still well-supported degrees, while the S\&P~500 profile largely levels off.
Consistent with this tail-persistence interpretation, the exploratory Fisher/BH analysis in Sec.~\ref{sec:supp_hvg_degree} places the first consecutive two-degree crypto--equity run at $k=12$ for Bitcoin and XRP and at $k=14$ for Ethereum.

At five and 10 minutes, the S\&P~500 also shows a pronounced rise over the supported moderate-degree range, and at 10 minutes it reaches the largest displayed $A(k)$ values around $k=6$--10.
The initial rise should therefore be read as a common localization of directional imbalance rather than as evidence of a persistent class separation.
Coarser sampling also maps each graph node to a longer physical interval and rapidly reduces support in the high-degree tail, so curves at a common $k$ do not represent an identical clock-time horizon or equal precision.
In contrast to the changing $D_\mathrm{JS}$ separation, the tail-restricted $A_\mathrm{KL}$ sign remains positive for all three cryptocurrencies and negative for the S\&P~500 at 1, 5, and 10 minutes.
Because $A_\mathrm{KL}$ summarizes normalized orientation rather than mismatch magnitude, this qualitative sign stability does not imply that the strength of irreversibility is preserved under aggregation.

\subsection{Complexity--Entropy Causality Plane}
\label{subsec:supp_cecp_sampling}

The same price reconstruction, aggregation rule, and matched 100-Shuffle/100-IAAFT ensembles were applied to the CECP diagnostic ($D=5$, $\tau=1$) over the same 1-, 5-, and 10-minute range as the DHVG check above, so that the two diagnostics are compared on identical sampling intervals.

\begin{figure}[pos=htbp]
    \centering
    \includegraphics[width=\textwidth]{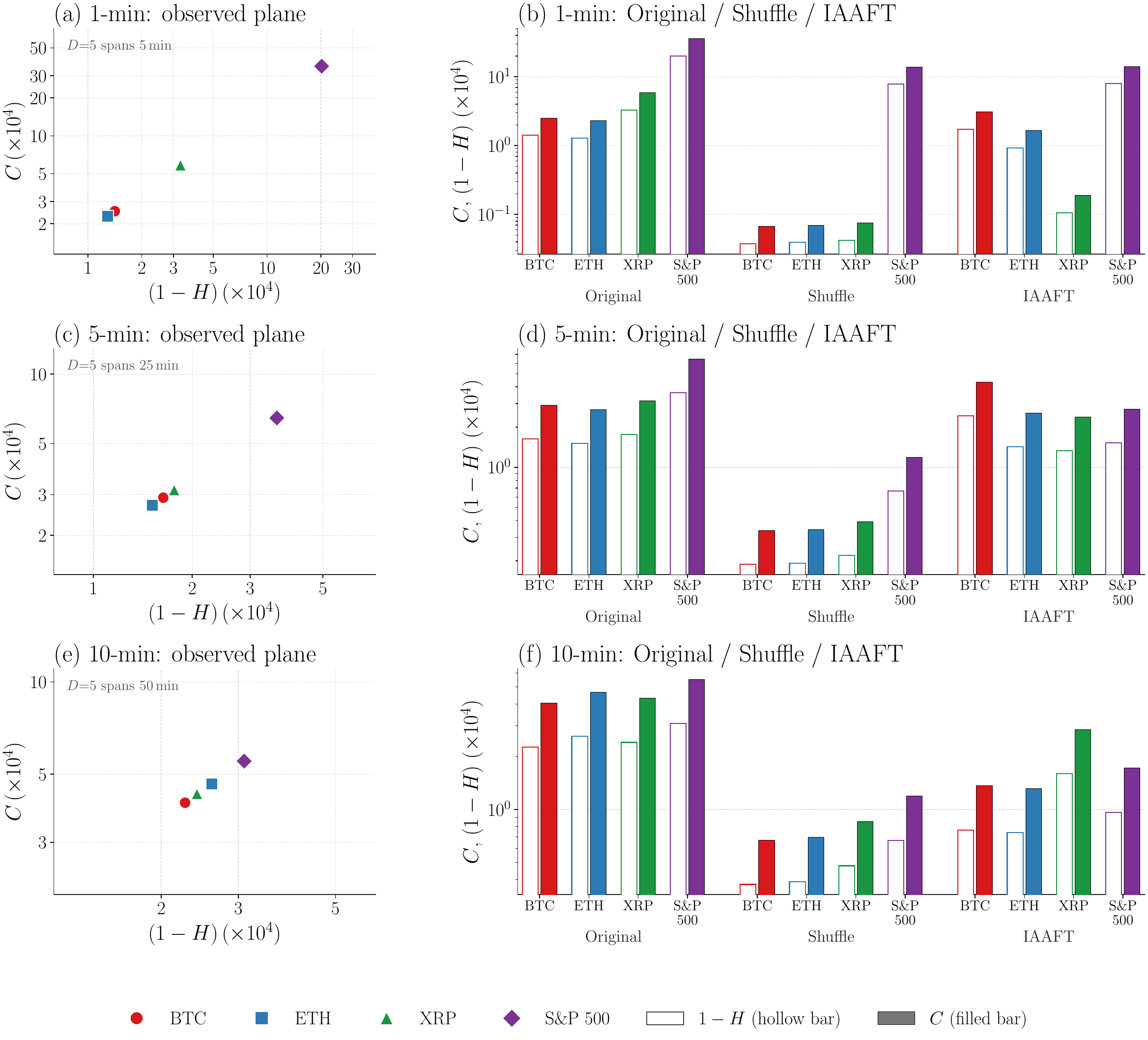}
    \caption{
        Sampling-interval sensitivity of the CECP diagnostic, in the layout of the main-text CECP figure.
        Rows show 1-, 5-, and 10-minute returns.
        Left column: observed $(1-H, C)$ plane (log--log).
        Right column: $10^4(1-H)$ (hollow bars) and $10^4 C$ (filled bars) for the Original, Shuffle, and IAAFT ensembles.
    }
    \label{fig:supp_cecp_sampling}
\end{figure}

    \begin{table}[htbp]
        \centering
        \caption{
            Combined sampling-interval sensitivity of the high-degree ($k\geq10$) source-gap DHVG and CECP diagnostics.
            For each sampling interval and asset, the Observed row gives the metric values.
            The second row reports the corresponding unadjusted inclusive Monte Carlo $p$-values separately for the Shuffle and IAAFT null models, in the order $p_{\rm Shuffle}/p_{\rm IAAFT}$.
            The $D_{\rm JS}$ comparison is upper-tailed, whereas the $A_{\rm KL}$, $(1-H)$, and $C$ comparisons are two-sided.
            The DHVG $D_{\rm JS}$ values are scaled by $10^2$, and the CECP coordinates are scaled by $10^4$.
            The displayed $p$-values are unadjusted within this 48-comparison interval family.
        }
        \label{tab:sampling_interval_combined}
        \setlength{\tabcolsep}{10pt}
        \renewcommand{\arraystretch}{1.2}
        {\footnotesize
        \begin{tabular}{@{}cc@{\hspace{10pt}}l@{\hspace{10pt}}rrrr@{\hspace{5pt}}}
            \toprule
            Interval & Asset & Entry & $10^2D_{\rm JS}^{k\geq10}$ & $A_{\rm KL}^{k\geq10}$ & $10^4(1-H)$ & $10^4C$ \\
            \midrule
            \multirow{2}{*}{1 min} & \multirow{2}{*}{BTC} & Observed & 0.2876 & +0.1238 & 1.413 & 2.517 \\
             &  & $p_{\rm Shuffle}/p_{\rm IAAFT}$ & 0.0100/0.0100 & 0.0400/0.0790 & 0.0200/0.0200 & 0.0200/0.0200 \\
            \cmidrule(lr){3-7}
            \multirow{2}{*}{1 min} & \multirow{2}{*}{ETH} & Observed & 0.2540 & +0.0653 & 1.287 & 2.292 \\
             &  & $p_{\rm Shuffle}/p_{\rm IAAFT}$ & 0.0100/0.0100 & 0.2570/0.1980 & 0.0200/0.0200 & 0.0200/0.0200 \\
            \cmidrule(lr){3-7}
            \multirow{2}{*}{1 min} & \multirow{2}{*}{XRP} & Observed & 0.3843 & +0.1106 & 4.925 & 8.744 \\
             &  & $p_{\rm Shuffle}/p_{\rm IAAFT}$ & 0.0100/0.0100 & 0.0990/0.0400 & 0.0200/0.0200 & 0.0200/0.0200 \\
            \cmidrule(lr){3-7}
            \multirow{2}{*}{1 min} & \multirow{2}{*}{S\&P 500} & Observed & 0.1017 & -0.0830 & 21.061 & 37.260 \\
             &  & $p_{\rm Shuffle}/p_{\rm IAAFT}$ & 0.1390/0.1090 & 0.2570/0.2970 & 0.0200/0.0200 & 0.0200/0.0200 \\
            \midrule
            \multirow{2}{*}{5 min} & \multirow{2}{*}{BTC} & Observed & 0.5793 & +0.1509 & 1.632 & 2.914 \\
             &  & $p_{\rm Shuffle}/p_{\rm IAAFT}$ & 0.0100/0.0100 & 0.0590/0.0400 & 0.0200/0.0200 & 0.0200/0.0200 \\
            \cmidrule(lr){3-7}
            \multirow{2}{*}{5 min} & \multirow{2}{*}{ETH} & Observed & 0.4161 & +0.2299 & 1.526 & 2.725 \\
             &  & $p_{\rm Shuffle}/p_{\rm IAAFT}$ & 0.0100/0.0300 & 0.0400/0.0200 & 0.0200/0.7130 & 0.0200/0.7130 \\
            \cmidrule(lr){3-7}
            \multirow{2}{*}{5 min} & \multirow{2}{*}{XRP} & Observed & 0.4073 & +0.1313 & 2.255 & 4.029 \\
             &  & $p_{\rm Shuffle}/p_{\rm IAAFT}$ & 0.0200/0.0100 & 0.0990/0.1190 & 0.0200/0.0200 & 0.0200/0.0200 \\
            \cmidrule(lr){3-7}
            \multirow{2}{*}{5 min} & \multirow{2}{*}{S\&P 500} & Observed & 0.4688 & -0.1602 & 3.724 & 6.647 \\
             &  & $p_{\rm Shuffle}/p_{\rm IAAFT}$ & 0.3370/0.3270 & 0.1190/0.1390 & 0.0200/0.0200 & 0.0200/0.0200 \\
            \midrule
            \multirow{2}{*}{10 min} & \multirow{2}{*}{BTC} & Observed & 0.3780 & +0.1881 & 2.269 & 4.041 \\
             &  & $p_{\rm Shuffle}/p_{\rm IAAFT}$ & 0.3270/0.3470 & 0.0400/0.0200 & 0.0200/0.0200 & 0.0200/0.0200 \\
            \cmidrule(lr){3-7}
            \multirow{2}{*}{10 min} & \multirow{2}{*}{ETH} & Observed & 0.4694 & +0.1568 & 2.635 & 4.694 \\
             &  & $p_{\rm Shuffle}/p_{\rm IAAFT}$ & 0.1190/0.1680 & 0.1390/0.0590 & 0.0200/0.0200 & 0.0200/0.0200 \\
            \cmidrule(lr){3-7}
            \multirow{2}{*}{10 min} & \multirow{2}{*}{XRP} & Observed & 0.5359 & +0.1389 & 2.451 & 4.394 \\
             &  & $p_{\rm Shuffle}/p_{\rm IAAFT}$ & 0.0790/0.0890 & 0.0590/0.1390 & 0.0200/0.0200 & 0.0200/0.0200 \\
            \cmidrule(lr){3-7}
            \multirow{2}{*}{10 min} & \multirow{2}{*}{S\&P 500} & Observed & 0.7135 & -0.0708 & 3.239 & 5.780 \\
             &  & $p_{\rm Shuffle}/p_{\rm IAAFT}$ & 0.5150/0.5840 & 0.5350/0.5350 & 0.0200/0.0200 & 0.0200/0.0200 \\
            \bottomrule
        \end{tabular}
        }
    \end{table}

Taken together, the CECP and DHVG results show a common sampling-interval dependence.
At one minute, the high-degree DHVG $D_\mathrm{JS}$ comparison separates all three cryptocurrencies from their IAAFT ensembles, while the CECP comparisons generally show clear original--IAAFT separation, most strongly for XRP and the S\&P~500.
BTC and ETH remain closer to their IAAFT ensembles in absolute CECP terms, consistent with the main-text linear-correlation interpretation.
As the interval increases from one to 10 minutes, the separation from the matched null ensembles contracts and the asset-dependent contrast becomes less resolvable across both diagnostics.
For the DHVG, the high-degree IAAFT $D_\mathrm{JS}$ comparison remains for all three cryptocurrencies at five minutes but clears no asset at 10 minutes.
For the CECP, the original--IAAFT gaps converge toward comparable relative sizes by 10 minutes, so the contrast visible at one minute is no longer distinguishable as an asset-specific pattern.

This shared weakening is consistent with aggregation mixing short-horizon fluctuations into longer return increments rather than with a stable aggregation-invariant effect.
Coarser aggregation also changes the effective sample size, lengthens the physical horizon spanned by each embedding window or degree bin, and, for the DHVG, thins high-degree tail support.
In particular, the $D=5$ CECP pattern spans 50 minutes of physical time at the 10-minute interval.
Sampling-interval robustness is therefore an open qualification shared by CECP and DHVG, rather than a property established independently for either diagnostic.

\bibliographystyle{elsarticle-num-names}
\bibliography{references}